\documentclass[preprint,10pt]{elsarticle}

\usepackage{amssymb}
\usepackage{amsmath}
\usepackage[caption=false,font=normalsize,labelfont=sf,textfont=sf]{subfig}
\usepackage{url}
\usepackage{hyperref}
\usepackage{algorithmic}
\usepackage{array}
\usepackage{multirow}
\usepackage{booktabs}
\usepackage{tabularx}
\usepackage{pgfplots}
\pgfplotsset{compat=1.12}
\usetikzlibrary{patterns}

\makeatletter
\def\ps@pprintTitle{%
 \let\@oddhead\@empty
 \let\@evenhead\@empty
 \let\@oddfoot\@empty
 \let\@evenfoot\@empty}
\makeatother

\makeatletter
\def\@affilmark#1{}
\makeatother



\begin{document}

\begin{frontmatter}

\title{Leveraging Broadcast Media Subtitle Transcripts for Automatic Speech Recognition and Subtitling}

\author{Jakob Poncelet}
\author{Hugo {Van hamme}}
\affiliation{organization={Department Electrical Engineering ESAT-PSI, KU Leuven}, country={Belgium}}

\begin{abstract}
The recent advancement of speech recognition technology has been driven by large-scale datasets and attention-based architectures, but many challenges still remain, especially for low-resource languages and dialects. This paper explores the integration of weakly supervised transcripts from TV subtitles into automatic speech recognition (ASR) systems, aiming to improve both verbatim transcriptions and automatically generated subtitles. To this end, verbatim data and subtitles are regarded as different domains or languages, due to their distinct characteristics. We propose and compare several end-to-end architectures that are designed to jointly model both modalities with separate or shared encoders and decoders. The proposed methods are able to jointly generate a verbatim transcription and a subtitle. Evaluation on Flemish (Belgian Dutch) demonstrates that a model with cascaded encoders and separate decoders allows to represent the differences between the two data types most efficiently while improving on both domains. Despite differences in domain and linguistic variations, combining verbatim transcripts with subtitle data leads to notable ASR improvements without the need for extensive preprocessing. Additionally, experiments with a large-scale subtitle dataset show the scalability of the proposed approach. The methods not only improve ASR accuracy but also generate subtitles that closely match standard written text, offering several potential applications.
\end{abstract}


\end{frontmatter}

\section{Introduction}
\noindent Speech recognition has seen remarkable improvements in recent years due to the introduction of large-scale datasets and attention-based architectures to the pre-training of speech models \cite{vaswani, conformer, bigssl, zhang2022pushing}. Massive multilingual pre-trained models are bridging the performance gap between high and low-resource languages \cite{xlsr}, although there are still many languages and local dialects for which the automatic transcriptions are far from human-quality \cite{feng2024}. Moreover, large-scale pre-training comes at a high cost that is often infeasible for small to medium scale businesses and academic researchers \cite{chen23l_interspeech}.

Most cutting-edge works in language and speech processing over the last decade have focused on self-supervised pre-training techniques to learn self-informed representations from large amounts of unlabelled data \cite{Chung2021w2vBERTCC, pmlr-v162-chiu22a}. Self-supervised learning (SSL) aims to train a speech encoder to extract informative features from raw audio without labels, based on a specific objective, e.g. predicting a masked out part of a spoken sentence while looking at the surrounding context \cite{ssl_review}. Popular models like Wav2vec 2.0 \cite{baevski2020wav2vec}, HuBERT \cite{hsu2021hubert}, WavLM \cite{wavlm} and their multilingual counterparts \cite{conneau21_interspeech, mhubert, wavlablm} have accomplished impressive improvements on many speech processing tasks, including Automatic Speech Recognition (ASR), requiring only a small amount of labelled data for task- or language-specific fine-tuning \cite{superb, mlsuperb}. 
Furthermore, pre-training these speech models on large quantities of diverse data improves generalisation to different domains \cite{hsu21_interspeech}.

\begin{figure*}[!t]
\centering
\includegraphics[width=\textwidth]{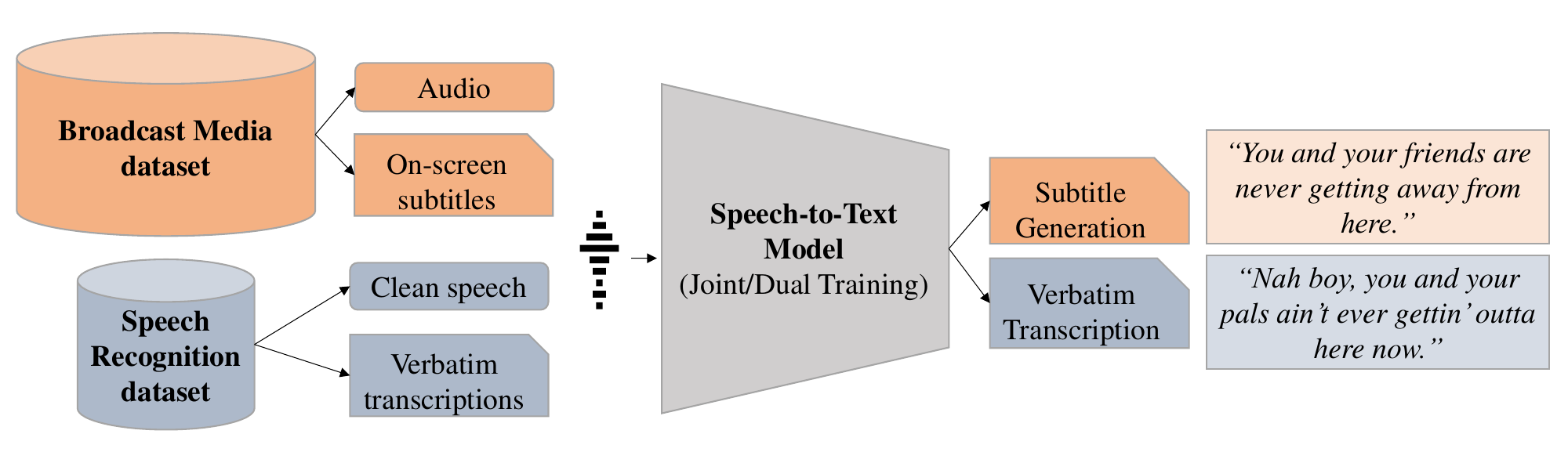}
\caption{Overview of the proposed approach. Verbatim transcriptions from ASR datasets and subtitle transcripts from a large source of broadcast media data are gathered. A dual speech-to-text model is trained to output a verbatim transcription for the input speech and to generate a well-suited subtitle at the same time, by jointly learning from both data streams.}
\label{fig:scheme}
\end{figure*}

Another strand of research in speech recognition has focused on building strong autoregressive decoders, by scaling up supervised training of ASR models with large amounts of data, either scraped from the web as in Whisper \cite{whisper} or collected from many datasets \cite{speechstew, owsm}. A powerful decoder can be trained to perform several speech processing tasks jointly, e.g. speech recognition, speech translation, language identification, voice activity detection and segmentation \cite{whisper}. Similar as in the unlabelled scenario, drastically scaling up the data sources makes the speech recognition models more robust \cite{likhomanenko21_interspeech}. Nevertheless, when ASR models are trained on weakly supervised web-scraped data as in Whisper, the advantages of manually labelled ASR datasets are often omitted, and the ability to transcribe exactly \textit{ad verbatim} is usually lost. Therefore, this work investigates the possibilities to include both manual transcripts from ASR datasets and weakly supervised transcripts in the form of TV subtitles into an ASR system, by jointly performing verbatim speech recognition and automatic subtitling in a multitask model, as illustrated in Figure~\ref{fig:scheme}.

TV subtitles are interesting for ASR training for several reasons. 
First, they are abundantly available in many languages and have been manually produced by human annotators.\footnote{In Flanders (Belgium), broadcasted media is subtitled specifically to support hard-of-hearing people, a practice legally mandated to some extent by the Flemish government, following European Union accessibility guidelines.}
Subtitlers apply specific rephrasings (e.g. shortening, paraphrasing), corrections (e.g. word choice, grammar, hesitations, repetitions) and normalisations (e.g. dialect) adhering to strong guidelines to improve the readability and comprehension for the viewer \cite{cintas2014audiovisual, MustCinema}. As such, subtitles offer a very useful mapping between spoken language, with all its disfluencies and mistakes, and more standardised written language. Second, subtitles cover many domains, ranging from read and prepared speech in broadcast news reports to conversational and spontaneous speech in talk shows, interviews, and even soaps and sitcoms. These real-life dialogues are far from the speech in most labelled speech datasets, like the read audiobooks in LibriSpeech \cite{librispeech}. Third, as a broader domain is covered, dialectal speech\footnote{In recent years, there has been a rise in the popularity of strongly dialectal speech in several TV shows in Flanders.}, accented speech and non-native speech are better represented than in common speech datasets. As the diversity of acoustic conditions, speech content and speaker effects is much higher in a large multimedia corpus, it has the potential to improve the robustness and generalisation of ASR models. Finally, as the amount of spoken multimedia content is growing at a fast pace, automatic subtitling solutions are becoming increasingly worthwhile and cost-effective, especially in globalised and international digital platforms.
Without any assistance, trained human subtitle transcribers require on average 13 to 18 times more time than the audio duration to generate a transcription \cite{roy09b_interspeech}. An automatic subtitling model already in the style of real subtitles would drastically reduce the effort for human subtitlers compared to a verbatim model. Moreover, transcripts in the style of standard written text can be more effectively utilised by Large Language Models (LLM), trained on large text datasets, compared to verbatim transcripts.

However, applying subtitles to ASR is non-trivial. As there is a big shift between exact, verbatim  transcriptions and standardised, clean TV subtitles, they are ineffective for direct E2E ASR training. On top of that, the timestamps of subtitles can be inaccurate and misaligned to the audio \cite{reazonspeech}.

In this work, we compare and propose several \textbf{supervised strategies} to combine verbatim and subtitle annotations to build strong multitask encoder-decoder ASR models, which improve verbatim speech recognition compared to standalone end-to-end (E2E) models and introduce enhanced subtitling capabilities. Our approach explicitly separates the subtitle and verbatim annotations to train a joint model and does not require extensive preprocessing (e.g. alignment, filtering, iterative refinement) of the subtitle data. There is no parallel data available for which both annotation types are present.

In previous work \cite{SubtitleModel}, we have proposed to combine both data streams with a parallel model, where the encoder is shared for both tasks, and two separate decoders are trained, conditioned on the outputs of the shared encoder. Although this strategy has the benefits of multitask learning and results in an improved encoder, it leaves out possibilities to improve the decoder and address the domain shift between the two datasets. First, we propose to view a subtitle as a translated version of the verbatim transcription. To this end, we add an additional subtitle encoder block and cascade it to the ASR encoder, to transform the verbatim output features of the ASR encoder into subtitle features. The two separate decoders are then conditioned on these different encoder outputs. As the encoded features of both encoders are useful for the decoders, we introduce a double cross-attention in the decoders to both encoder outputs. We investigate several architectures that incorporate these ideas. Second, we explore a task-specific shared decoder that is conditioned on a task token representing which transcription type to generate, as has been proposed for multilingual \cite{owsm} and multitask \cite{ihori23_interspeech} training.

All experiments are carried out on Belgian Dutch (Flemish), which is a medium-resourced language with many local dialects and a significant shift between written language and colloquial spoken language. We start from a strong Conformer-based, encoder-decoder E2E ASR baseline to which we make the proposed adaptations. All models are evaluated for verbatim transcription and subtitling capabilities on test sets with read speech and prepared talks and with spontaneous speech. We also carry out scaling experiments, showing the benefits of large-scale data, and introduce serialised output training \cite{kanda20b_interspeech}. While the direct E2E ASR model degrades by using the subtitles as if they were verbatim, the proposed models show strong improvements on all evaluations. Finally, we compare our method to Whisper and evaluate a pipeline approach that generates subtitles with an LLM from ASR transcripts.

The main \textbf{contributions} of this research paper can be summarised as:
\vspace{-\topsep}
\begin{itemize}
    \itemsep-4pt
    \item The investigation of several approaches to combine subtitles with verbatim data for end-to-end ASR training,
    \item Proposing new architectures and supervised training techniques specifically tailored for weakly-supervised transcripts, more specifically TV subtitles, which strongly improve ASR performance,
    \item The expansion of a joint modelling framework to construct models that are able to generate both a verbatim transcription and a subtitle, and
    \item An extensive evaluation of different training paradigms on small and large-scale datasets, and a comparison to state-of-the-art models and methods.
\end{itemize}
All produced code and resulting models are made open-source.\footnote{Code: \url{https://github.com/nelfproject/NeLF_Transcription_ASR}}\footnote{Models: \url{https://huggingface.co/nelfproject}}

\section{Related Work}   \label{sec:relatedwork}
\noindent In this work, we use subtitles as a valuable resource to improve the accuracy and robustness of an ASR system. This section gives an overview of recent advancements within this field and related fields that make use of subtitles. The proposed subtitle method is also situated within the context of weakly supervised ASR training with imperfect labels. In some works, generating a transcription in the same language as is spoken (\textit{intralingual}) is termed captioning, while generating a transcription in a different language (\textit{interlingual}) is termed subtitling \cite{xu-etal-2022-joint}. This work focuses only on intralingual subtitling. We use the term subtitling, as all data has arisen from subtitles on TV, and captions can have a broader meaning (e.g. image captioning).

\subsubsection{Subtitles in ASR}
\noindent Traditionally, there have been two main methods to leverage subtitle data for ASR. First, when generating pseudo-labels of broadcast media data with a pre-trained acoustic model, the subtitles can be exploited to filter and/or refine bad hypotheses based on some alignment metric, and then the generated transcripts are used to iteratively refine the acoustic model \cite{lamel_2002, lanchantin16_interspeech, BangIEICE}. Similarly, the subtitles themselves can also be used as training targets to gradually build an acoustic model on a larger corpus by iteratively refining the alignment \cite{Ando2021ConstructionOA, reazonspeech}, which can be done with external models \cite{bell15_ASRU, Saz2018, JHU_kaldi} or sophisticated preprocessing algorithms \cite{BangIEICE}. Second, subtitles can be used to refine ASR outputs, either by training a biased (often program-based or genre dependent) language model on the subtitles \cite{lamel_2002, BangIEICE, vishwa2015}, or with postprocessing techniques, e.g. to restore punctuation \cite{Guerreiro21eswa, geislinger-etal-2022-improved, Milde2021_1109} or compress the transcript for optimised screen readability \cite{liu-etal-2020-adapting}. Several of these works have been inspired by or have resulted from challenges like the Multi-Genre Broadcast (MGB) challenge \cite{MGB1, MGB2, iberspeech}.

\subsubsection{Subtitles in Speech Translation}
\noindent Most production-house movies and series are subtitled concurrently in many languages, leading to a big manual corpus suited for multilingual machine and speech translation \cite{MustCinema}. Recent work \cite{xu-etal-2022-joint} has investigated a dual decoder model to simultaneously produce an intralingual and a translated subtitle from the output of a pre-trained verbatim ASR model. To improve over general cascaded models \cite{che2017}, end-to-end speech translation models have also been proposed that directly produce a translated subtitle from speech, e.g. by regularising the encoder with a CTC loss for the intralingual subtitle and generating a translated subtitle with the decoder \cite{papi-etal-2023-direct-speech}. In that case, alignment of the subtitle in the source language can be generated with a segmentation algorithm on the CTC outputs \cite{kurzinger2020}, and these timings can subsequently be mapped to alignments in the target language \cite{papi-etal-2023-direct-speech}.

\subsubsection{Written Text Generation}
\noindent A similar strand of research considers the differences between written text and spoken text transcriptions \cite{ihori-etal-2020-parallel, liao23, ihori23_interspeech}. In that case, spoken text is the verbatim transcription, and written text is an adapted version where disfluencies and fillers are removed and punctuation marks are added, which can then be used for Natural Language Processing (NLP) applications \cite{nozaki22_interspeech, futami23_icassp}. While subtitle generation is close to the task of written text transcription, there are many effects like rephrasings, summarisations and harsh subphrase deletions that are very specific to subtitles and not part of these written text annotations. Therefore, subtitles are a weaker form of supervision than written text, which is basically filtered and punctuated verbatim text. While many systems rely on post-hoc processing of the output of an ASR model for written text generation, using e.g. inverse text normalisation \cite{sunkara21_icassp}, disfluency and filler removal \cite{wang-etal-2022-adaptive}, and spelling correction \cite{guo19_icassp}, an end-to-end model that jointly learns to transcribe spoken and written text has been proposed which outperforms separate models and the cascaded approach \cite{ihori23_interspeech}, by leveraging a shared decoder that is conditioned on the task to perform.

\subsubsection{Weakly Supervised ASR}
\noindent There has been a long-standing interest in improving acoustic modelling for ASR using weak supervision \cite{li2017, li2019, lamel_2002}. Typically, the weak labels are used to filter or improve the outputs of a pre-trained acoustic model, such that they can be used for ASR training \cite{lamel_2002}. Recent work has shown that end-to-end speech recognition models can be trained with incomplete or partial reference labels \cite{pratap2022_neurips}, unordered reference labels \cite{pratap2022_icassp} or even contextually related labels (e.g. from social media video captions) \cite{sing2020_icassp}. The popularity of weakly supervised training has mainly been driven by the rise of very large datasets \cite{chen21o_interspeech}, which are often created by web-scraping audio sources and forced-aligning transcriptions extracted from the web or ASR outputs \cite{wenetspeech, galvez2021peoples, reazonspeech}. Finally, recent work has shown that competitive and robust ASR models can be built by training on a huge corpus of web-scraped supervised speech data as in Whisper \cite{whisper}, which can be considered weakly labelled, although various filtering techniques and data curation methods are in order.

\subsubsection{Proposed Approach}
\noindent This work differs from previous efforts by jointly utilising the verbatim transcriptions from a standard ASR dataset and the subtitle transcriptions from a large dataset of broadcast media data, and carefully designing a model that can learn from and also generate both modalities. The model is trained completely end-to-end and the method does not require any preprocessing (e.g. filtering, pseudo-labeling or forced alignment) of the weakly labelled data, nor any postprocessing (e.g. an external program-based LM, inverse text normalisation), nor an iterative refinement of the acoustic models and/or data, although some of these techniques could still be applied. Finally, in this work, we do not focus on predicting line breaks and screen breaks \cite{wilken-etal-2022-suber}, which can be done with segmentation models \cite{karakanta-etal-2022-evaluating}.

\section{Methods} \label{sec:methods}
\noindent We propose several architectures to improve end-to-end verbatim ASR using subtitled data. We start from a strong Conformer-based encoder-decoder ASR baseline. The following section details all proposed models, their differences and the integration of both verbatim and subtitled data.

\begin{figure*}[h!]
    \centering
    \subfloat[c][Naive E2E ASR - Shared Decoder \label{fig:baseline}]{\includegraphics[width=0.45\textwidth]{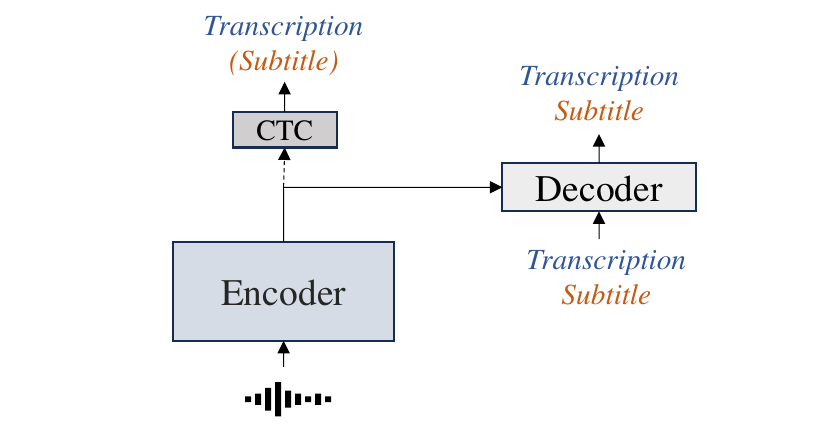}}
    \hfill
    \subfloat[c][Shared Task Decoder \label{fig:taskdecoder}]{\includegraphics[width=0.45\textwidth]{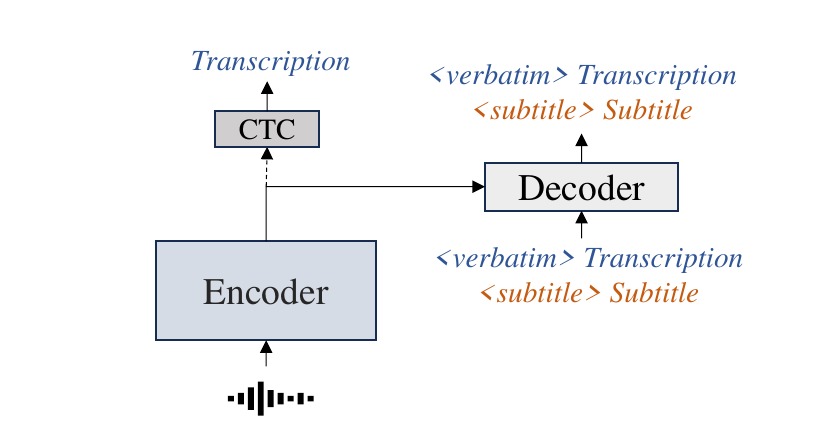}}
    
    \subfloat[c][Parallel Decoders \label{fig:parallel}]{\includegraphics[width=0.45\textwidth]{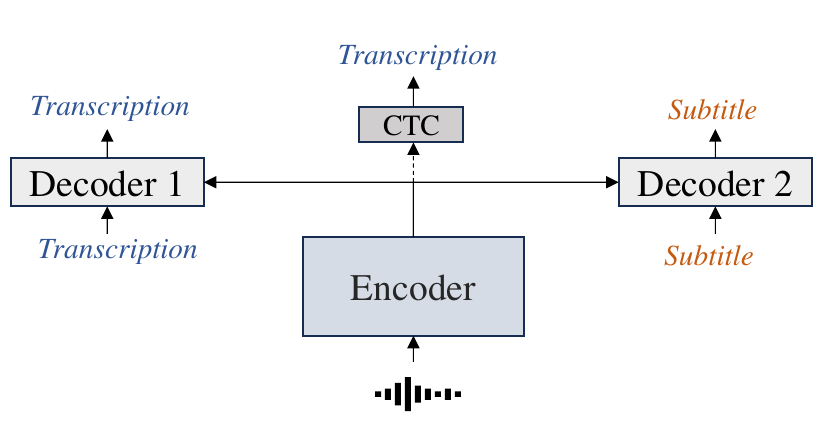}}
    \hfill
    \subfloat[c][Cascaded Encoder Features \label{fig:cascadeenc}]{\includegraphics[width=0.45\textwidth]{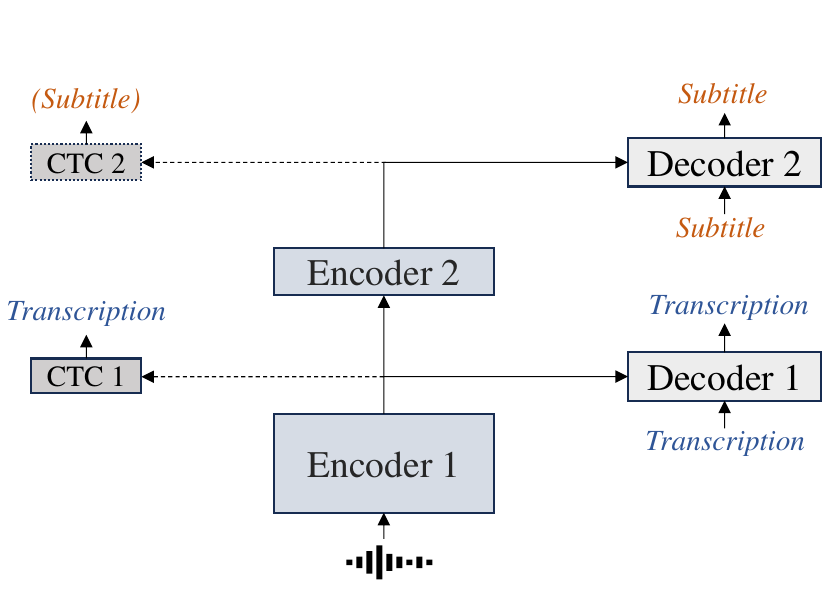}}
    
    \medskip
    \subfloat[c][Cascaded Decoder Features \label{fig:cascadedec}]{\includegraphics[width=0.45\textwidth]{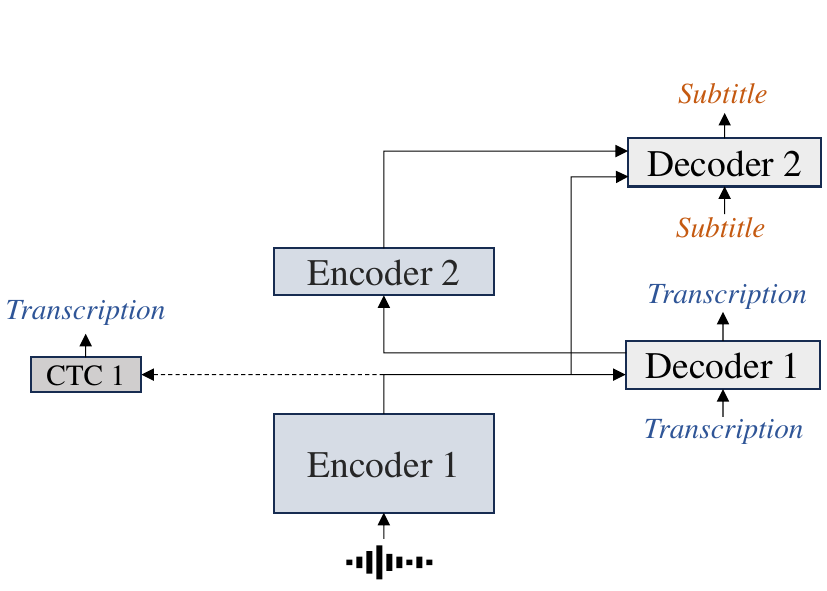}}
    \hfill
    \subfloat[c][Cascaded Encoder Dual Features \label{fig:cascadedual}]{\includegraphics[width=0.45\textwidth]{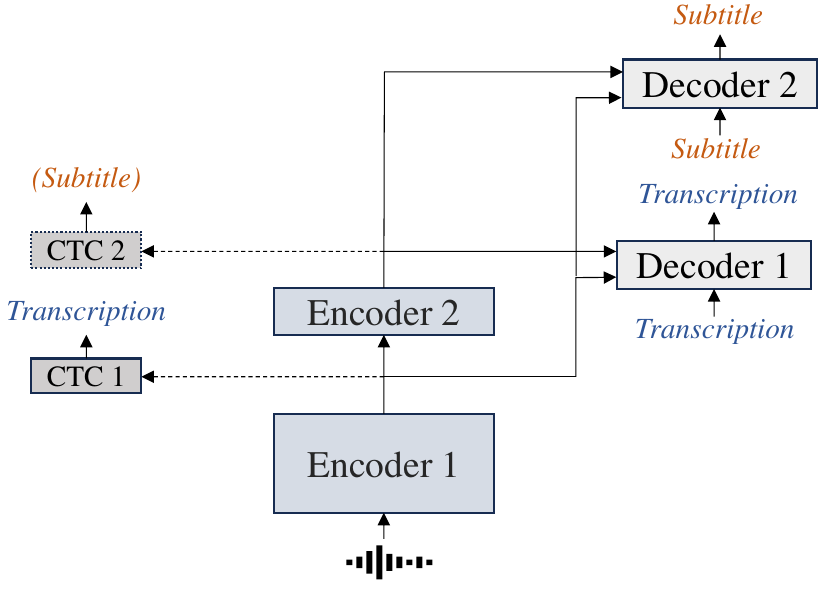}}
    \caption[Model comparison]{Comparison of all proposed models.}
    
    \smallskip
    \scriptsize
    \begin{tabularx}{\textwidth}{l@{  } X}
    \hline 
    (a) & E2E ASR: Encoder-decoder ASR model with CTC regularisation. In \textit{naive} E2E ASR, the subtitles are treated as verbatim transcriptions and both datasets are combined. \\
    (b) & Shared task decoder model: similar to (a), but the decoder is conditioned on a task token to generate either a verbatim or a subtitle output. \\
    (c) & Parallel model: the encoder is shared and there are two separate decoders, one for verbatim transcription (Decoder~1) and one for subtitling (Decoder~2). \\
    (d) & Cascaded encoder model: the subtitle encoder (Encoder~2) processes the outputs of the (shared) ASR encoder (Encoder~1). The verbatim decoder (Decoder~1) is conditioned on the ASR encoder outputs, and the subtitle decoder (Decoder~2) on the outputs of both encoders.\\
    (e) & Cascaded decoder model: similar to (d), but the subtitle encoder uses as input the last layer states of the ASR decoder instead of the ASR encoder outputs. \\
    (f) & Cascaded model using dual encoder features: similar to (d), but the verbatim ASR decoder is conditioned on the outputs of both the verbatim and the subtitle encoder.\\
    \hline
    \end{tabularx}
    \label{fig:fig_sim}
\end{figure*}

\subsection{End-to-end ASR}
\noindent In end-to-end ASR, the raw audio inputs are directly mapped to a transcription. It is customary to apply an encoder-decoder model to this sequence-to-sequence problem. The encoder extracts some informative features from the acoustic input, after which the decoder builds a sentence autoregressively, conditioned on the acoustic evidence in the encoder states. 

In this work, the ASR encoder is a Conformer \cite{conformer}, which combines global self-attention with local convolutions to compute informative feature representations. To regularise the encoder and guide it towards acoustically meaningful outputs, a Connectionist Temporal Classification (CTC) \cite{ctc} objective is imposed on its final outputs and on intermediate layer outputs \cite{interctc}. CTC predicts a monotonic alignment between the encoder features and the output sequence, but assumes that the states are independent conditioned on the data. 

A Transformer decoder \cite{vaswani} is trained to predict the output sequence autoregressively, by relying on self-attention to its previously predicted tokens and a cross-attention to the encoder outputs \cite{watanabe2017}. The decoder can attend to the encoder’s predictions to generate a better alignment and build a transcription, token by token. In fact, the decoder is trained to classify the correct output token given the previous tokens in that utterance, like a language model. 

The encoder-decoder hybrid CTC/Attention model \cite{watanabe2017} is optimised end-to-end, with a weighted sum of both objectives. The loss function in Eq.~\ref{eq:asrloss} is computed as a sum of the CTC regularisation loss (on the encoder's output $\mathcal{L}_{ctc}$ and an intermediate layer $\mathcal{L}_{interctc}$) and the cross-entropy classification loss $\mathcal{L}_{att, asr}$ of the decoder, weighted by the regularisation parameters $\alpha$ (CTC weight) and $\beta$ (inter-CTC weight).
\begin{equation}
	\mathcal{L}_{asr} = (1 - \alpha) \mathcal{L}_{att, asr} + \alpha \left[ (1-\beta) \mathcal{L}_{ctc} + \beta \mathcal{L}_{interctc} \right]
 \label{eq:asrloss}
\end{equation}

To prevent overconfidence of the model during training, the decoder class targets are softened using label smoothing \cite{smoothing}. During decoding, the model combines the CTC and attention-based output probabilities in a joint beam search. CTC pre-scores the hypothesis tokens to inform the attention decoder of the expected alignment and remove irrelevant hypotheses.

A naive way to incorporate subtitles into the baseline E2E ASR model, is to treat the subtitles as regular verbatim annotations and directly use them standalone or mixed with verbatim ASR data. We call this approach \textit{naive end-to-end ASR}. It is shown in Figure \ref{fig:baseline}.

\subsection{Parallel Model}
\noindent As subtitles are not exact transcripts, they can degrade the verbatim output predictions of the decoder when naively combining the verbatim transcripts and the subtitles. A model with separate decoders for both tasks solves this issue, as we proposed in \cite{SubtitleModel}. Both decoders attend to the encoder’s output, but one is trained to generate a verbatim ASR transcription, while the other is trained to generate a subtitle transcription. The decoders work independently in parallel. By keeping the encoder shared, the model still enjoys the advantage of processing both data streams, and thus is adapted to both domains in a multi-task fashion \cite{tied2018}. Since the CTC objective assumes a monotonic alignment, and subtitles align very differently (due to e.g. rephrasing, compression and summarisation), the CTC loss is only incurred on the verbatim data. 

During training, both data types are mixed and the network optimizes a weighted combination of the verbatim ASR loss and the subtitling loss. The ASR loss $\mathcal{L}_{asr}$ from Eq.~\ref{eq:asrloss} is backpropagated over the verbatim data only, masking the subtitle utterances in the verbatim decoder. The subtitle loss contains the subtitle decoder’s attention loss $\mathcal{L}_{att, subs}$ and is backpropagated over the subtitle data (masking the verbatim utterances). The parameters $\lambda_{asr}$ and $\lambda_{subs}$ can be tuned to weigh the importance of both tasks during optimisation. The final loss is given in Eq.~\ref{eq:joint}. The parallel model is depicted in Figure~\ref{fig:parallel}.
\begin{equation}
	\mathcal{L}_{tot} = \lambda_{asr} \mathcal{L}_{asr} + \lambda_{subs} \mathcal{L}_{att, subs}
    \label{eq:joint}
\end{equation}

\subsection{Cascaded Model}
\noindent As argued in the previous section, there is a mismatch between verbatim and subtitle annotations. Therefore, using a completely shared encoder could limit its capabilities, as both objectives might counteract each other when attended to by the decoders. To remedy this, we introduce an additional (smaller) encoder, which we call the subtitle encoder, that should act as a translation block from verbatim to subtitle. The subtitle encoder, consisting of a few Transformer encoder layers, is cascaded to the outputs of the ASR encoder. The cascaded model is inspired by advances in end-to-end speech translation, where the model is decomposed in multiple encoder-decoder structures, each optimised for a given task and chained together \cite{fastmd, dalmia-etal-2021-searchable}. We will refer to the shared encoder (like the one in the parallel and naive methods) as the ASR encoder, since its output is used for the verbatim CTC objective and the verbatim ASR decoder. The subtitle encoder computes the features for the subtitle decoder. We investigate three possibilities for the cascaded model, which all incorporate the concept of a Multi-Transformer decoder.

\subsubsection{Multi-Transformer}
\noindent We define a Multi-Transformer decoder as a Transformer block with multi-sequence attention, as proposed in previous work for speech translation \cite{dalmia-etal-2021-searchable} and multimodal translation \cite{helcl-etal-2018-cuni}. In a regular Transformer encoder-decoder model, there is one (multi-head) cross-attention layer in every decoder layer that attends to the encoder states. The Multi-Transformer decoder has two cross-attention layers, attending to different encoders, per decoder layer to leverage multiple information streams.

\subsubsection{Cascaded Encoder Features} 
\noindent The first approach for a cascaded model is depicted in Figure~\ref{fig:cascadeenc}. In the first scheme, the outputs of the ASR encoder are directly fed as inputs to the subtitle encoder. Notice that the ASR encoder is regularised with CTC using verbatim labels (for verbatim data only). Then, the subtitle encoder transforms the acoustic verbatim features towards subtitle-like features. Finally, to not lose any temporal information in the subtitle encoder block, the subtitle decoder is augmented to a Multi-Transformer decoder. It can attend to the ASR encoder output and to the subtitle encoder output, effectively combining both streams to produce its best subtitle hypothesis.

Because of the translating subtitle encoder block, it is also possible to apply an additional CTC loss to the output of the subtitle encoder for subtitle data, although we found only limited benefits in doing this. Previous work \cite{chuang-etal-2021-investigating, yan-etal-2023-ctc} has argued that a Transformer model with a CTC objective is able translate an input sequence to a non-monotonic alignment, despite the monotonicity assumptions in the CTC framework (e.g. in machine translation). Applying a subtitle CTC loss was not possible in the previous methods, as a shared CTC module for verbatim and subtitled data, or separate CTC objectives applied to the same encoder output, would lead to contradictory objectives for the encoder layers due to the different alignments between verbatim transcriptions and subtitles. Still, even without an additional CTC objective, the subtitle encoder block allows the model to translate verbatim features to subtitle features, without hampering the verbatim CTC and the verbatim ASR decoder objective.

\subsubsection{Cascaded Decoder Features}
\noindent In the second scheme, shown in Figure~\ref{fig:cascadedec}, the final hidden states of the ASR decoder are used as input for the subtitle encoder. For subtitle data, there is no verbatim reference transcription such that backpropagation from the ASR decoder outputs is not possible during training. One possible solution would be to run the ASR decoder in inference mode (i.e. perform beam search) for subtitle data during training. However, this drastically slows down training since the parallelism is lost, and does not align well with the joint training approach proposed here. Another solution is to generate verbatim pseudo-labels for the subtitled data with a pre-trained ASR model (or with the current model after a certain number of epochs) and forward them through the ASR decoder. However, this is also not very scalable and is very slow in case of large subtitle datasets. 

Based on these considerations, we use an alternative approach. If there is no reference verbatim transcription for the utterance (i.e. for subtitle utterances), we forward an ${<}unk{>}$ token through the decoder as verbatim reference (which is masked in the ASR decoder's loss) and let the verbatim decoder generate a sentence embedding for this token. We observed that this improves the optimisation of the ASR decoder. As during training all temporal acoustic detail is lost when using the decoder states as input, the subtitle decoder is again converted to a Multi-Transformer attending to both encoders’ outputs, so that it can still access the temporal information in the ASR encoder. As such, the subtitle decoder can leverage the verbatim sentence embedding produced by the ASR decoder as well as the acoustic features produced by the ASR encoder to generate a subtitle prediction.\footnote{We note that there is a mismatch between training and testing in terms of decoder states, but the subtitle decoder learns to focus on the first ASR decoder's hidden state (i.e. the sentence embedding). This approach has more potential in case iterative re-training with generated verbatim pseudo-labels is applied.} Due to the sequence length reduction when going to verbatim labels, the subtitle CTC loss is not applied here.

\subsubsection{Cascaded Encoder Dual Features}
\noindent The third scheme, depicted in Figure~\ref{fig:cascadedual}, is similar to the first scheme, directly cascading the encoder blocks. Intuitively, the verbatim decoder might benefit from a proposed subtitle to generate a transcription. In this method, both decoders are Multi-Transformer decoders and attend to the two encoders’ outputs. The amount of subtitle data is larger than the amount of verbatim data, but it’s not directly backpropagated through the ASR decoder. Hence, the ASR decoder can now attend to the subtitle encoder, which is trained on all subtitle data, via the additional cross-attention layer.

\subsubsection{Loss Function} 
\noindent The computation of the loss in the cascaded models is similar to the parallel model. The only difference is the possible inclusion of the subtitle-specific CTC loss $\mathcal{L}_{ctc, subs}$ with weight $\gamma$. The final objective function is detailed in Eq.~\ref{eq:cascaded}.
\begin{equation}
	\mathcal{L}_{tot} = \lambda_{asr} \mathcal{L}_{asr} + \lambda_{subs} \left[ (1 - \gamma) \mathcal{L}_{att, subs} + \gamma \mathcal{L}_{ctc, subs} \right]
    \label{eq:cascaded}
\end{equation}

\subsection{Shared Task Decoder}
\noindent The previous methods propose separate decoders to solve the verbatim ASR and subtitling task, due to the mismatch in domains. Inspired by recent advances in multi-task ASR decoder training \cite{owsm}, we extend the decoder to be able to perform both tasks while sharing the weights, by prepending a task token to the transcriptions, as proposed in \cite{ihori23_interspeech} for dual spoken and written text transcription. For verbatim ASR data, the token ${<}verbatim{>}$ is added before the start-of-sentence token in the transcription. For subtitle data, the token ${<}subtitle{>}$ is used. In this setup, both data streams can be combined without confounding the decoded sequences. The decoder conditions its hypothesis on the task token and can deliver a verbatim or a subtitle type transcription depending on the given task token. The advantage of this model is that the decoder is optimised with all of the data, although it has to do more heavy lifting. Figure~\ref{fig:taskdecoder} presents the proposed method.

\section{Experimental Setup}  \label{sec:setup}
\noindent This section describes the details of the setup used in the experiments, including the model configurations, training aspects and the datasets.

\subsection{Datasets}  \label{sec:data}
\noindent This paper proposes several methods to combine a verbatim annotated ASR dataset with a large collection of subtitled audio in the same language, i.e. Belgian Dutch (Flemish). 

\subsubsection{Verbatim data}
\noindent As a source of verbatim data, we use a standard speech recognition database called Corpus Gesproken Nederlands (CGN) or Spoken Dutch Corpus \cite{CGN_Oostdijk}. The Flemish part of the dataset contains 270 hours of manually annotated speech recordings, divided over a multitude of components which each represent a specific type of speech and environment. Among the components, we can find 1) prepared, read speech by professional (news)readers, 2) recordings of interviews, lectures and meetings, 3) live sports commentary, 4) narrowband telephone conversations and 5) face-to-face discussions and conversational speech. We have derived a training set of 240 hours of speech (350k utterances), which we call \textit{cgn-train}. For the scaling experiments, where the ratio of subtitle data to verbatim data is increased, we use a 3-fold speed perturbed version of \textit{cgn-train} with speed perturbation factors 0.9, 1.0 and 1.1 \cite{ko15_interspeech}. Furthermore, we use a representative test set of 8 hours of speech (7k utterances, 83k words) by sampling several recordings from every component, excluding telephone recordings, with a complete separation of speakers between train and test set. This test set, called \textit{cgn-dev}, corresponds with previous work \cite{ponceletASRU, SubtitleModel}. Finally, we also created a verbatim test set \textit{subs-annot} of 6 hours from broadcast TV data, which is described in the next subsection.

\subsubsection{Subtitled data}
\noindent We have compiled a large in-house dataset of subtitled data as broadcasted on Flemish TV. The subtitles and corresponding audio streams have been provided by the Flemish public broadcaster VRT. Due to copyright restrictions, this data cannot be released open-source. For the experiments, we use three data splits. The first split, used for analysis of the proposed models, contains 720 hours of speech (915k utterances), identical to \cite{SubtitleModel}. This split is a collection of audio streams from several topics, including TV talk shows with many different guests, broadcast TV news, live interviews, soap series, political talk shows, etc. For every recording, we have the corresponding subtitle as it appeared on screen. We have built a subtitle dataset by segmenting the audio recordings based on these screen timings. However, the screen timings are far from perfect, which makes the supervision a lot weaker as well. The subtitles have been normalised to some extent by filtering out music and non-speech (e.g. applause, ringtones) if they were annotated consistently (e.g. music is often tagged with an asterisk, non-speech audio events are reported in all uppercase). In a second stage, this dataset has been scaled up to 2000 hours with recordings from the same or similar sources. This second split will be used to perform initial scaling experiments. In a final stage, we have built a dataset of 14000 hours of audio recordings with all kinds of content, either broadcasted on TV or online. Because of the scale of this third and final data split, we only do a few experiments on this dataset. The three training splits will be referred to as \textit{subs-720h}, \textit{subs-2kh} and \textit{subs-14kh} respectively. For subtitle evaluation, we retain two held-out data splits. The first one originates from the 2000 hours of data, which we refer to as \textit{subs-valid}, and consists of 11 hours (110k words) of speech. The second one is a sample from the entire 14000 hours dataset, which we call \textit{subs-valid-14kh}, and consists of 22 hours (190k words) of subtitled data.

Lastly, we have created a verbatim test set called \textit{subs-annot}, where we selected a representative subset of the TV shows in the \textit{subs-720h} set, gathered new data from these shows and manually transcribed it \textit{ad verbatim}. The \textit{subs-annot} test set contains 6 hours of transcribed speech (5k utterances, 71k words) and corresponds with previous work \cite{SubtitleModel}.

\subsection{Model details}

\subsubsection{Input format}
\noindent The audio recordings are converted to 16~kHz wav-files. We extract 80-dimensional mel-filterbanks and 3-dimensional pitch features using a window of 25~ms and a frame shift of 10~ms. The filterbank and pitch features are concatenated and mean-variance normalised at the utterance level. During training, SpecAugment \cite{park19e_interspeech} augmentations are applied to the input features, adaptively masking spans of time windows and frequency bands and warping time frames.

\subsubsection{Model layout} \label{sec:model}
\noindent All proposed models are modifications of the baseline encoder-decoder ASR model, implemented in ESPnet \cite{watanabe18_interspeech}. The ASR encoder is a Conformer \cite{conformer}, preceded by a Conv2d input layer which subsamples the input features four-fold and transforms them to the hidden dimension of 256. Relative positional encodings are added to the inputs and used in the self-attention Conformer layers. The Conformer encoder has 12 layers in macaron-style with Swish activations. Every layer has 4 attention heads, the feed-forward layer dimension is 2048 and the CNN has a kernel size of 31. The decoders are 6-layer Transformer decoders, with 4 attention heads, 2048 feed-forward units and the same hidden dimension as the encoder. Dropout of 0.1 is applied at all layers. These hyperparameters arise from standard recipes for ASR models \cite{watanabe18_interspeech} and have been deducted based on optimal performance of the baseline ASR model. In case the model implements an additional subtitle encoder, it consists of 2 regular Transformer layers with the same dimensions as the ASR Conformer encoder layers. For the experiments on the \textit{subs-2kh} and \textit{subs-14kh} data, the subtitle encoder consists of 6 Transformer layers. For the scaling experiments in Section~\ref{sec:scaling}, we train a larger variant, denoted \textit{XL-model}, with a hidden dimension of 512 and 8 attention heads per layer. The base variants contain 70M parameters and the XL variant has 180M parameters. The baseline ASR model has 50M parameters.

\subsubsection{Training} 
\noindent The models are trained with a joint hybrid CTC/Attention loss \cite{watanabe2017}, with CTC weight $\alpha = 0.3$ applied to the encoder outputs for the verbatim targets only. Intermediate CTC \cite{interctc} is applied to layer 6 of the ASR encoder, with $\beta = 0.3$. When indicated, subtitle CTC loss is applied with $\gamma = 0.3$. During multitask training, a batch consists of an equal amount of utterances from the verbatim and subtitle datasets. If the subtitle dataset is larger than the verbatim dataset, the verbatim utterances are oversampled in an epoch (except in the naive method). The transcripts are tokenized into Byte-Pair Encoding (BPE) (sub)word units. The BPE model has a vocabulary of 5000 unigram BPE’s and is trained on both the verbatim transcriptions and an equally-sized sample of the subtitles, such that the verbatim and subtitle transcriptions can share the same BPE token space.
All transcripts are normalised and lower-cased without punctuation. The decoder predicts the target BPE’s which are smoothed with a label smoothing weight of 0.1. By default, the prediction losses of the ASR encoder and the subtitle decoder are weighted with an equal weight $\lambda_{asr} = \lambda_{subs} = 0.5$, which was found optimal in previous work \cite{SubtitleModel} when mixing batches equally with utterances from both data streams.

The models are trained on a single GPU (20GB), with an effective batch size of 1024 and the Adam optimiser \cite{adam} with an exponential decaying learning rate with a peak of 0.004 and 25k linear warmup steps. The models are trained until convergence with a maximum of 150 epochs for \textit{subs-720h} and 30 epochs for \textit{subs-2kh}. The 10 best intermediate training checkpoints with the highest validation accuracy are averaged for evaluation. For the large-scale experiments on \textit{subs-14kh}, the models are trained for 5 epochs with a learning rate of 0.001 and 100k warmup steps on 16 GPU's (40GB) by accumulating gradients over GPU's.

\subsubsection{Decoding}
\noindent During inference, the models generate a verbatim transcription and a subtitle in parallel. The decoder integrates the CTC prefix-scores with a weight of 0.3. The best hypotheses for both decoders are computed with a beam search keeping the 20 best running hypotheses. No language model is applied during decoding.

\subsection{Evaluation}
\noindent The verbatim transcription outputs of the models are evaluated based on the Word Error Rate (WER) with respect to the reference verbatim transcripts. The WER metric is modified to take into account equally correct transcriptions in Belgian Dutch, e.g. for filler words, hyphenations, number normalisation and words with multiple correct spellings. The subtitle outputs are evaluated based on a BLEU score \cite{bleu, sacrebleu} with respect to the real subtitles on screen, such that differences in sentence ordering are punished less. We use a smoothed BLEU-4 score with uniform weights for all n-grams ($n=1..4$).

The statistical significance of the results is analysed by computing the \textit{p}-values between hypotheses from different models. For ASR results (WER), the MAPSSWE \cite{mapsswe} test from the NIST Scoring Toolkit (SCTK)\footnote{\url{https://github.com/usnistgov/SCTK}} is used. For subtitling results (BLEU), we conduct paired tests with bootstrap resampling ($n=1000$ fold) \cite{koehn-2004-statistical} with the SacreBLEU\footnote{\url{https://github.com/mjpost/sacrebleu}} toolkit. All results from different models (on the same test set) that are \textit{statistically insignificant} from each other with $p>0.05$ are marked with the same tag ($\mathbf{\dagger}$ or $\mathbf{\ddagger}$) in tables. If more than two results in a table have the same tag, all pairwise comparisons are also statistically insignificant.

\section{Experiments}   \label{sec:exp}
\noindent In this section, we perform several experiments and compare the proposed methods to established baseline models. Both the verbatim transcription and subtitle generation capabilities of the models are assessed on different benchmarks. Furthermore, we analyse the effect of the double cross-attention in the Multi-Transformer Decoder layers, the different output modalities of the verbatim and subtitle decoder, and the effect of data filtering based on the reference subtitles. Moreover, we perform a scaling experiment on a large subtitle dataset using long-form punctuated transcriptions. Finally, we compare our proposed model to the related state-of-the-art Whisper \cite{whisper} speech model. We also evaluate whether LLMs can be used to generate subtitles from ASR outputs and compare this approach to our end-to-end model.

\subsection{Performance Comparison}

\subsubsection{Verbatim Speech Recognition}
\noindent First, the proposed models are evaluated on a verbatim speech recognition task. All models are trained on two subtitle datasets, consisting of 720 hours (\textit{subs-720h}) and 2000 hours (\textit{subs-2kh}) of speech respectively, combined with the verbatim CGN dataset (\textit{cgn-train}). All models are evaluated on \textit{cgn-dev} and \textit{subs-annot}, by comparing the outputs of the verbatim ASR decoders to the reference verbatim transcriptions. Results are shown in Table~\ref{tab:asr1}.

Naively using subtitles as verbatim transcriptions leads to a degradation on the ASR dataset \textit{cgn-dev}, and even diverges when the subtitle-to-verbatim ratio is too high. As the proposed models are able to distinguish between the two domains, they don’t suffer from these drawbacks. The parallel decoder model benefits from domain adaptation in the shared encoder yielding substantial gains on the \textit{subs-annot} set, which is adopted from TV series, but lacks strong improvements on \textit{cgn-dev} with less data. The shared task-decoder model shows nice improvements on \textit{cgn-dev}, benefiting from the additional data to train the decoder. However, it performs worse on \textit{subs-annot}. It is likely that the decoder has learnt the difference between the two training datasets, and therefore outputs more subtitle-like transcripts for the \textit{subs-annot} set, despite being conditioned on the task token. This is a phenomenon also reported by other works involving multi-task decoders \cite{owsm}. Finally, the cascaded models report the lowest WERs, in both data regimes. The cascaded encoder model is essentially the same as the parallel model, except for the additional subtitle encoder block, which gives more freedom to the model to encode the verbatim and subtitle targets differently without hampering the CTC objective applied to the ASR encoder. The cascaded model with dual encoder features performs best on the \textit{subs-annot} broadcast TV test set, as the verbatim decoder is able to attend to both the ASR encoder and the subtitle encoder outputs. The cascaded models result in a 20 to 30\% improvement on the verbatim speech recognition task, without any additional verbatim data or filtering methods. In addition, they produce a dual subtitle output, evaluated in the next section.

\begin{table}
\caption{WERs (\%) of verbatim speech recognition experiments ($\downarrow$). The models are trained using the verbatim data from CGN and either 720h or 2000h subtitle data. They are evaluated on \textit{cgn-dev} and \textit{subs-annot}.}
\footnotesize
\label{tab:asr1}
\begin{center}
\begin{tabular}{@{\extracolsep{2pt}} l | c  c  c  c @{}}
\toprule
\hfill \textit{Train} & \multicolumn{2}{c}{\textit{720h subtitles}} & \multicolumn{2}{c}{\textit{2000h subtitles}} \\
\cline{2-3} \cline{4-5} \noalign{\vskip 1mm}
\textbf{Model} \hfill \textit{Test}& \textit{cgn-dev}~ & \textit{subs-annot}~ & \textit{cgn-dev}~ & \textit{subs-annot}~ \\
\midrule
E2E ASR - CGN only & 10.71~ & 14.06~ & 10.71~ & 14.06~ \\
Naive E2E ASR & 13.74~ & 30.34~ & 97.67~ & 94.17~ \\
Shared Task Decoder & 9.38~ & 20.99~ & 8.51~ & 16.75~ \\
Parallel Decoders & 10.01~ & 10.76~ & 8.84~ & 10.07~ \\
Cascaded Encoder & \textbf{8.78}~ & 9.93$^\dagger$ & 8.27$^\dagger$ & 9.65~ \\
Cascaded Decoder & 8.94$^\dagger$ & 10.09$^\dagger$ & \textbf{8.26}$^\dagger$ & 9.33$^\dagger$ \\
Cascaded Enc. Dual & 8.99$^\dagger$ & \textbf{9.89}$^\dagger$ & 8.29$^\dagger$ & \textbf{9.26}$^\dagger$ \\
\bottomrule 
\end{tabular}
\end{center}
\end{table}

\subsubsection{Automatic Subtitling}
\noindent Second, the subtitle outputs of the models are evaluated with a BLEU score with respect to the reference on-screen subtitle. To this end, for cascaded and parallel models, the output of the subtitle decoder is used. For the task-specific model, the decoder is conditioned on the subtitle task token. These are the same models as in Table~\ref{tab:asr1}, trained using either 720 hours or 2000 hours of subtitled data. All multitask models are evaluated on \textit{subs-valid}, and the BLEU scores are reported in Table~\ref{tab:subs1}.

The parallel model is not able to generate strong subtitles, compared to the shared task decoder and the cascaded models. In most cases, the cascaded models produce the highest quality subtitles with very promising BLEU scores. The cascaded encoder model slightly has the edge over the other architectures.

\begin{table}
\caption{BLEU scores (\%) of subtitle recognition experiments ($\uparrow$). The models are trained using the verbatim data from CGN and either 720h or 2000h subtitle data. They are evaluated on \textit{subs-valid}. For the baseline ASR model and the naive E2E ASR model, the predictions of the ASR decoder are scored.}
\footnotesize
\begin{center}
\label{tab:subs1}
\begin{tabular}{@{\extracolsep{2pt}} l | c c @{}}
\toprule
\hfill \textit{Train} & \textit{720h subtitles} & \textit{2000h subtitles} \\
\cline{2-2} \cline{3-3} \noalign{\vskip 1mm}
\textbf{Model} \hfill \textit{Test} & \textit{subs-valid}~ & \textit{subs-valid}~ \\
\midrule
E2E ASR - CGN only & 29.88~ & 29.88~ \\ 
Naive E2E ASR & 51.05~ & 33.32~ \\
Shared Task Decoder & 50.63$^\dagger$ & 52.56~  \\
Parallel Decoders & 46.16~ & 46.35~ \\
Cascaded Encoder & \textbf{51.17}~ & \textbf{54.14}$^\dagger$ \\
Cascaded Decoder & 50.67$^\dagger$ & 53.95$^\dagger$ \\
Cascaded Enc. Dual & 50.58$^\dagger$ & 54.06$^\dagger$ \\
\bottomrule 
\end{tabular}
\end{center}
\end{table}

\subsection{Analysis and Ablation}

\subsubsection{Multi-Transformer Decoder}
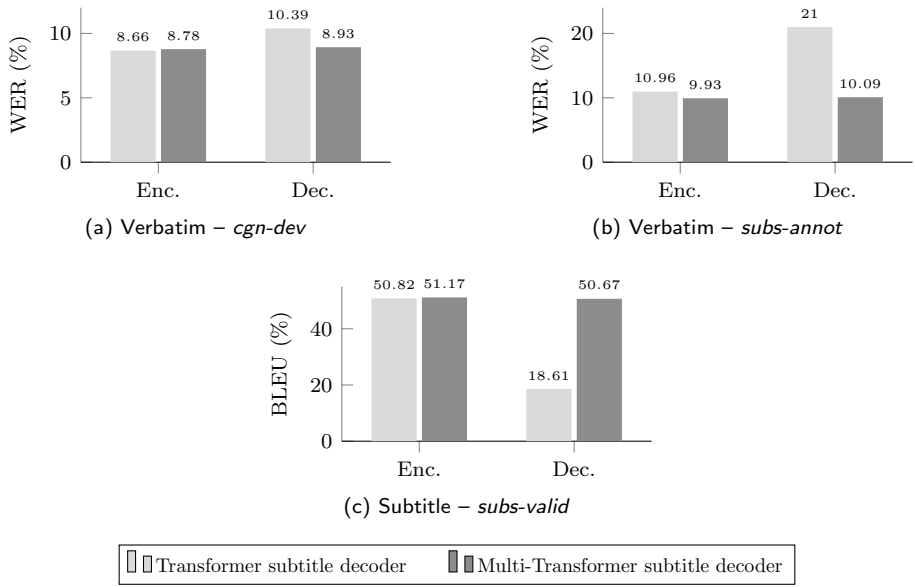
\begin{figure}[t]
\captionsetup[subfigure]{font=footnotesize}
\centering
\subfloat[c][Verbatim -- \textit{cgn-dev}]{%
\begin{tikzpicture}
\begin{axis}[
        ybar, width=0.47\textwidth, height=0.3\columnwidth,
        bar width=0.6cm, ymin=0, ymax=12,
        axis x line*=bottom, axis y line*=left,
        y axis line style={opacity=0.5},
	ylabel={WER (\%)}, ylabel style={align=center, font=\footnotesize},
        yticklabel style={font=\footnotesize},
        xticklabel style={font=\footnotesize},
        enlarge x limits=0.5,
        legend entries={Transformer subtitle decoder, Multi-Transformer subtitle decoder},
        legend style={
            legend columns=-1, 
            /tikz/every even column/.append style={column sep=0.5cm},
            nodes={scale=0.75},
            font=\normalsize
        },
        legend to name=named_legend,
        symbolic x coords={Enc., Dec.}, 
        xtick=data,
        nodes near coords,
        every node near coord/.append style={font=\tiny},
        nodes near coords align={vertical},
]%
\addplot [draw=none,fill=gray!30] coordinates {
  (Enc.,8.66) (Dec.,10.39)
};
\addplot [draw=none,fill=gray!90] coordinates {
  (Enc.,8.78) (Dec.,8.93)
};
\end{axis}
\end{tikzpicture}%
}
\hfill
\subfloat[c][Verbatim -- \textit{subs-annot}]{%
\begin{tikzpicture}
\begin{axis}[
        ybar, width=0.47\textwidth, height=0.3\columnwidth,
        bar width=0.6cm, ymin=0, ymax=24,
        axis x line*=bottom, axis y line*=left,
        y axis line style={opacity=0.5},
	ylabel={WER (\%)}, ylabel style={align=center, font=\footnotesize},
        yticklabel style={font=\footnotesize},
        xticklabel style={font=\footnotesize},
        enlarge x limits=0.5,
        symbolic x coords={Enc.,Dec.}, xtick=data,
        nodes near coords,
        every node near coord/.append style={font=\tiny},
        nodes near coords align={vertical},
]
\addplot [draw=none,fill=gray!30] coordinates {
  (Enc.,10.96) (Dec.,21.00)};
\addplot [draw=none,fill=gray!90] coordinates {
  (Enc.,9.93) (Dec.,10.09)};
\end{axis}
\end{tikzpicture}%
}

\subfloat[c][Subtitle -- \textit{subs-valid}]{%
\begin{tikzpicture}
\begin{axis}[
        ybar, width=0.47\columnwidth, height=0.3\columnwidth,
        bar width=0.6cm, ymin=0, ymax=55,
        axis x line*=bottom, axis y line*=left,
        y axis line style={opacity=0.5},
	ylabel={BLEU (\%)}, ylabel style={align=center, font=\footnotesize},
        yticklabel style={font=\footnotesize},
        xticklabel style={font=\footnotesize},
        enlarge x limits=0.5,
        symbolic x coords={Enc.,Dec.}, xtick=data,
        nodes near coords,
        every node near coord/.append style={font=\tiny},
        nodes near coords align={vertical},
]
\addplot [draw=none,fill=gray!30] coordinates {
  (Enc.,50.82) (Dec.,18.61)};
\addplot [draw=none,fill=gray!90] coordinates {
  (Enc.,51.17) (Dec.,50.67)};
\end{axis}
\end{tikzpicture}%
}
\center{\ref{named_legend}}
\caption{Comparison between subtitle decoder blocks with only one cross-attention to the output of the subtitle encoder (``Transformer"), or with two cross-attentions, i.e. once to the output of the ASR encoder and once to the output of the subtitle encoder (``Multi-Transformer"). The subtitle encoder is either conditioned on the ASR encoder (``Enc.") or decoder (``Dec.") features.}
\label{fig:multitransf}
\end{figure}

\noindent We investigate the quantitative and qualitative effect of the proposed Multi-Transformer Decoder, i.e. subtitle decoder blocks with two consecutive cross-attention layers both attending to different inputs, once attending to the ASR encoder output and once to the subtitle encoder output. The cascaded encoder model (Fig.~\ref{fig:cascadeenc}) and cascaded decoder model (Fig.~\ref{fig:cascadedec}), which have a Multi-Transformer subtitle decoder, are compared to their respective variants with regular Transformer subtitle decoders without the additional cross-attention (i.e. in that case, the arrow from Encoder 1 to Decoder 2 is dropped in Fig.~\ref{fig:fig_sim}). In the cascaded encoder model, the subtitle encoder block is conditioned on the outputs of the ASR encoder, while in the cascaded decoder model the subtitle encoder is conditioned on the last layer of the ASR decoder. Figure \ref{fig:multitransf} depicts the WER and BLEU scores for this ablation study. All differences are statistically significant except the encoder results in Figure~\ref{fig:multitransf}a.

The additional cross-attention to the ASR encoder’s output in the Multi-Transformer decoder leads to an improvement over the single cross-attention Transformer decoder. For the cascaded decoder model, the ASR encoder outputs are necessary for fine-grained temporal detail which is lost in the decoder features that are fed to the subtitle encoder, leading to expected improvements. The cascaded encoder model enjoys notable improvements on \textit{subs-annot}, as the additional cross-attention leads to a stronger backpropagation through the ASR encoder for the subtitled data.

\subsubsection{Dual Outputs -- Translation effects} \label{sec:translation}
\noindent To understand why the subtitles differ so strongly from verbatim transcriptions, some additional information about the peculiarities of the data is required. In the last decades, the predominant spoken language in Flanders (the Northern, Dutch-speaking part of Belgium) has been changing towards a variant called ``\textit{tussentaal}'' or ``intermediate language'', which is structurally positioned in between the local Flemish dialects and the Dutch standard language \cite{vandekerckhove2009_subtitling, remael2008}. In contrast to Standard Belgian Dutch (SBD), the official written language, this variant is called Colloquial Belgian Dutch (CBD): an all-encompassing term for the multiple regiolects in Flanders, which are strongly subjected to regional and social variation \cite{prieels2018}. While Standard Dutch remains the norm in formal domains and non-fictional informational television (e.g. documentaries), informal speech is shifting from local dialects to regiolects \cite{vandekerckhove2009_dialect} and has become a public medium, omnipresent in less-informational TV programs \cite{vandekerckhove2009_subtitling}. CBD has several key features that differentiate it from SBD, which can mainly be categorised into lexical (e.g. ``\textit{appelsien}" instead of ``\textit{sinaasappel}" both meaning the fruit ``orange"), morphological (e.g. diminutive ``-\textit{ke}" instead of ``-\textit{je}", personal pronoun ``\textit{ge}/\textit{gij}" instead of ``\textit{je}/\textit{jij}" meaning ``you") and syntactic features (e.g. using a double negation like ``don't know nothing") \cite{vandekerckhove2007, prieels2018}. These could be compared to colloquialisms and slang in English (e.g. ``bloody knackered" instead of ``very exhausted" in British English, ``ain't" instead of ``isn't" in American English), as well as contractions used in informal speech (e.g. ``gonna/going to", ``wanna/want to", ``kinda/kind of" and ``innit/isn't it"). Generally, regiolectal speech is not transcribed literally on Flemish television, but according to guidelines converted into SBD written form \cite{remael2008}. Especially CBD features that are closer to dialect are more often converted to SBD in subtitles \cite{prieels2018}.

This linguistic discrepancy between spoken and written language is reflected in the differences between the outputs of the subtitle decoder and the verbatim decoder. The subtitle decoder maps the spoken dialectal language towards its Standard Dutch version, sometimes with complete rephrasing, while the verbatim decoder does not make these conversions. Furthermore, the subtitle decoder corrects hesitations, repetitions and other disfluencies, as well as typical abbreviations in spontaneous speech, which are less common in written text. In \ref{sec:app-ex}, we have included several examples that demonstrate these differences.

\subsubsection{Data Filtering}
\noindent Most efforts \cite{lamel_2002, lanchantin16_interspeech, BangIEICE} leverage subtitles to filter the hypotheses of a pre-trained ASR model, so that the remaining subtitles (or hypotheses) can be used as reference verbatim transcriptions to train an improved ASR model. In our approach, such filtering methods are not required, as the models can distinguish between verbatim and subtitle labels, but can still be applied complementary to the proposed training pipeline. To remove subtitles which are too different from the uttered sentence (e.g. misaligned with bad timings, harsh deletions, completely different), we perform some experiments with automated data filtering. To this end, a segment is removed if the verbatim prediction of a pre-trained ASR model is inconsistent with the on-screen subtitle. The pre-trained ASR model is the baseline from Table~\ref{tab:asr1}, trained on \textit{cgn-train}. The BLEU scores between the hypotheses and the subtitles are computed, and the segments with a BLEU score below the filtering threshold are dropped. A new ASR model (standard or cascaded) is then trained on the combination of the filtered dataset and \textit{cgn-train}. The effects of filtering on dataset size and WER are shown in Table~\ref{tab:filt}. A cascaded encoder model, trained with subtitle CTC loss and two Transformer decoders, is compared to the naive approach.

For the naive approach, where there is no explicit difference between subtitles and verbatim transcriptions, strongly filtering out divergent subtitles is beneficial for the model. For the cascaded model, some filtering can be useful to throw out the misaligned speech segments with incorrect timestamps. Data filtering in this case is not mandatory as the subtitles and verbatim transcriptions are nicely separated, and only useful feedback from the subtitles is backpropagated to improve the ASR modelling. Hence, our approach allows to conveniently scale up the weak supervision.

\begin{table}
\caption{Data filtering experiments for verbatim speech recognition, evaluated in terms of WER ($\downarrow$) on \textit{cgn-dev} and \textit{subs-annot}.}
\footnotesize
\label{tab:filt}
\begin{center}
\resizebox{\columnwidth}{!}{
\begin{tabular}{@{\extracolsep{2pt}} l | c  c  c | c c @{}}
\toprule
& Subtitle & Filtering & Remaining & \multicolumn{2}{c}{WER (\%)} \\
\textbf{Model} & dataset & BLEU & subtitles & \textit{cgn-dev}~ & \textit{subs-annot}~ \\
\midrule
\multirow{3}{*}{Naive E2E ASR} & \textit{none} & 0\% & 0 h & 10.71~ & 14.06~ \\
& \textit{subs-720h} & 0\% & 720 h & 13.74~ & 30.34~ \\
& \textit{subs-2kh} & 60\% & 270 h & 12.10~ & 11.11~ \\
\noalign{\vskip 1mm} \hline \noalign{\vskip 1mm}
\multirow{5}{*}{Cascaded Encoder} & \multirow{2}{*}{\textit{subs-720h}} & 0\% & 720 h & 8.80~ & 10.39~ \\
 &  & 10\% & 480 h & 9.37~ & 10.69~ \\
\noalign{\vskip 1mm} \cline{2-6} \noalign{\vskip 1mm}
 & \multirow{3}{*}{\textit{subs-2kh}} & 0\% & 2000 h & 8.90$^\dagger$ & 11.56~ \\
 & & 10\% & 1200 h & \textbf{8.74}$^\dagger$ & 11.16~ \\
 & & 60\% & 270 h & 9.43~ & 11.86~ \\
\bottomrule 
\end{tabular}
}
\end{center}
\end{table}

\subsection{Scaling Experiments}

\subsubsection{Long-form Punctuated Training}
\noindent To build a powerful model of high performance that is readily deployable for automatic transcription services, we extend the capabilities of the decoders by enriching the target transcriptions. To this end, we add complete punctuation and capitalisation instead of normalising the transcriptions (which was done in the previous experiments), so that the model can generate a fully formatted transcription. For the verbatim transcriptions, we also add transcription tags (e.g. ${<}{*}a{>}$, ${<}{*}v{>}$) to some words indicating when a word is cut off mid-word, when it is a foreign or dialect word, etc. Those tags are part of the rich manual annotations of the verbatim dataset \cite{CGN_Oostdijk}. 

Furthermore, we noticed that an utterance in the dataset is on average only 3 seconds long. Training a baseline model on these short utterances as in previous experiments (Table~\ref{tab:asr1}), leads to a Word Error Rate of 10.75 \% on the \textit{cgn-dev} test set of equally short utterances (4 seconds on average), but degrades to 15.12 \% when evaluating on the same test set but with concatenated utterances of 10 seconds on average. We denote this test set \textit{cgn-dev-long}. To remedy this effect, we combine short consecutive utterances in the training dataset, so that the utterance durations are more uniformly distributed with an average duration of 9 seconds and a maximum of 15 seconds. Both the verbatim and subtitle dataset are transformed into a long-form equivalent by concatenating utterances. Additionally, when two consecutive utterances come from different speakers, we add a speaker change token ${<}spk{>}$ between the utterances, following serialised output training techniques \cite{shafey19_interspeech, kanda20b_interspeech, kanda22_interspeech}. For the subtitle dataset, this is based on the colour of the subtitle (which should switch when there are different speakers). This pushes the model to learn acoustically when different people talk within one segment, without requiring corpus-level speaker information. Analogous to \textit{cgn-dev-long}, we create a long-form equivalent of \textit{subs-annot} (4 seconds on average) and denote it \textit{subs-annot-long} (9 seconds on average).

Table~\ref{tab:comb1} shows the results of a first experiment with long-form data and serialised output training. The models are trained on a subset of 20\% of the 2000 hours subtitle dataset from previous experiments (in order to reduce the computational load in these initial experiments) and evaluated on the short-form test sets \textit{cgn-dev} and \textit{subs-annot}, and their longer form equivalents \textit{cgn-dev-long} and \textit{subs-annot-long}. The first row shows the results of an ASR model trained on short-form and normalised data, the other results are from long-form models.
Additionally, Table~\ref{tab:comb2} shows the BLEU scores of the evaluated models on the short form \textit{subs-valid} set, and on the longer form \textit{subs-valid-14kh} set, which is also more difficult as it covers a wider range of TV shows and types of speech. The additional tags (verbatim tags, punctuation, speaker change) are neglected during WER and BLEU calculation.

The trends observed for long-form training are similar to previous training setups.
Note that compared to Table~\ref{tab:asr1} and Table~\ref{tab:subs1}, less subtitle data is used, and the decoder does a lot more (tag prediction, punctuation, etc.).
For verbatim ASR, the cascaded models outperform the other approaches, as in the short-form setting. The cascaded model with dual encoder features significantly outperforms the other methods for most test sets. For subtitling, in general the same trends across models are visible. The BLEU scores on the short-form \textit{subs-valid} in Table~\ref{tab:comb2} are lower than in previous experiments, due to less subtitled training data used and the short-form effects observed in Table~\ref{tab:comb1}. For practical use of ASR models, this long-form training method is more robust to long utterances and multi-speaker fragments, while generating a fully formatted output including punctuation, capitalisation and anomaly tagging.

\begin{table}
\begin{center}
\caption{WERs (\%) of verbatim speech recognition experiments ($\downarrow$) with long-form serialised output training. The models are evaluated on the short-form \textit{cgn-dev} and \textit{subs-annot}, and the long-form \textit{cgn-dev-long} and \textit{subs-annot-long}.}
\footnotesize
\label{tab:comb1}
\resizebox{\columnwidth}{!}{
\begin{tabular}{@{\extracolsep{2pt}} l | c  c  c  c @{}}
\toprule
Model & \textit{cgn-dev}~ & \textit{cgn-dev-long}~ & \textit{subs-annot}~ & \textit{subs-annot-long}~ \\
\midrule
E2E ASR - CGN only (S) & \underline{10.75}$^\dagger$ & \underline{15.12}~ & 14.06~ & 16.85$^\ddagger$ \\
E2E ASR - CGN only & 13.03$^\ddagger$ & 10.61~ & 15.05~ & 14.17~ \\
Shared Task Decoder & 13.34$^\ddagger$ & 12.32~ & 18.42~ & 17.86$^\ddagger$ \\
Parallel Decoders & 13.22$^\ddagger$ & 10.75~ & 13.00~ & 13.13~ \\
Cascaded Encoder & 11.09$^\dagger$ & 9.09$^\dagger$ & 11.59$^\dagger$ & 11.64$^\dagger$ \\
Cascaded Decoder & 11.14$^\dagger$ & 9.09$^\dagger$ & 11.30$^\dagger$ & 11.40$^\dagger$ \\
Cascaded Enc. Dual & \textbf{11.00}$^\dagger$ & \textbf{8.86}~ & \textbf{10.80}~ & \textbf{10.84}~ \\
\bottomrule 
\end{tabular}
}
\end{center}
\end{table}

\begin{table}
\begin{center}
\caption{BLEU scores (\%) of subtitle prediction experiments ($\uparrow$) with long-form serialised output training. The models are evaluated on \textit{subs-valid} and \textit{subs-valid-14kh}. For the E2E ASR model, the output of the (verbatim) ASR decoder is used. For the other models, the predictions of the subtitle decoder are used.\\}
\label{tab:comb2}
\footnotesize
\begin{tabular}{@{\extracolsep{2pt}} l | c  c @{}}
\toprule
Model & \textit{subs-valid}~ & \textit{subs-valid-14kh}~ \\
\midrule
E2E ASR - CGN only (S) & 29.88~ & 33.71~ \\
E2E ASR - CGN only & 32.01~ & 34.06~ \\
Shared Task Decoder & 39.27$^\dagger$ & \textbf{45.12}~ \\
Parallel Decoders & 35.51~ & 37.02~ \\
Cascade Encoder & \textbf{39.83}~ & 42.95~ \\
Cascade Decoder & 39.18$^\dagger$ & 43.36$^\dagger$ \\
Cascade Enc. Dual & 39.41$^\dagger$ & 43.51$^\dagger$ \\
\bottomrule 
\end{tabular}
\end{center}
\end{table}

\subsubsection{Large Scale Modelling}  \label{sec:scaling}
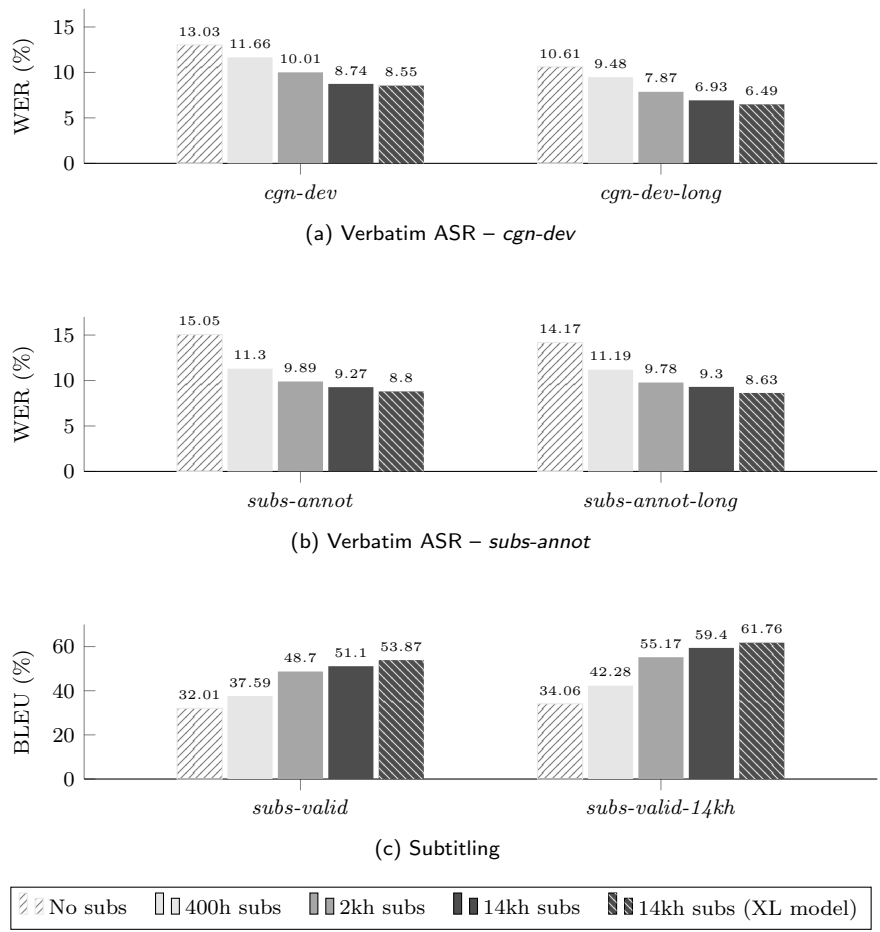
\begin{figure*}
\captionsetup[subfigure]{font=footnotesize}
\centering
\subfloat[c][Verbatim ASR -- \textit{cgn-dev}]{%
\begin{tikzpicture}
\begin{axis}[
        ybar, width=\textwidth, height=0.3\columnwidth,
        bar width=0.60cm, ymin=0, ymax=17,
        axis x line*=bottom, axis y line*=left,
        y axis line style={opacity=0.5},
	ylabel={WER (\%)}, ylabel style={align=center, font=\footnotesize},
        yticklabel style={font=\footnotesize},
        xticklabel style={font=\footnotesize},
        enlarge x limits=0.6,
        symbolic x coords={cgn-dev,cgn-dev-long,subs-annot,subs-annot-long},
        xtick=data,
        xticklabels={\textit{cgn-dev},\textit{cgn-dev-long}},
        legend entries={No subs, 400h subs, 2kh subs, 14kh subs, 14kh subs (XL model)},
        legend style={
                legend columns=-1,
                /tikz/every even column/.append style={column sep=0.3cm},
                font=\footnotesize
        },
        legend to name=named_legend_2,
        nodes near coords,
        every node near coord/.append style={font=\tiny},
        nodes near coords align={vertical},
]%
\addplot [draw=gray!20,fill=gray!20,pattern={north east lines},pattern color=gray] coordinates {(cgn-dev,13.03) (cgn-dev-long,10.61)};
\addplot [draw=none,fill=gray!20] coordinates {(cgn-dev,11.66) (cgn-dev-long,9.48)};
\addplot [draw=none,fill=gray!70] coordinates {(cgn-dev,10.01) (cgn-dev-long,7.87)};
\addplot [draw=none,fill=black!70] coordinates {(cgn-dev,8.74) (cgn-dev-long,6.93)};
\addplot [draw=none,fill=black!70,postaction={pattern=north west lines,pattern color=gray!30}] coordinates {(cgn-dev,8.55) (cgn-dev-long,6.49)};
\end{axis}
\end{tikzpicture}%
}

\bigskip
\subfloat[c][Verbatim ASR -- \textit{subs-annot}]{%
\begin{tikzpicture}
\begin{axis}[
        ybar, width=\textwidth, height=0.3\columnwidth,
        bar width=0.60cm, ymin=0, ymax=17,
        axis x line*=bottom, axis y line*=left,
        y axis line style={opacity=0.5},
	ylabel={WER (\%)}, ylabel style={align=center, font=\footnotesize},
        yticklabel style={font=\footnotesize},
        xticklabel style={font=\footnotesize},
        enlarge x limits=0.6,
        symbolic x coords={cgn-dev,cgn-dev-long,subs-annot,subs-annot-long},
        xtick=data,
        xticklabels={\textit{subs-annot},\textit{subs-annot-long}},
        nodes near coords,
        every node near coord/.append style={font=\tiny},
        nodes near coords align={vertical},
]%
\addplot [draw=gray!20,fill=gray!20,pattern={north east lines},pattern color=gray] coordinates {(subs-annot,15.05) (subs-annot-long,14.17)};
\addplot [draw=none,fill=gray!20] coordinates {(subs-annot,11.30) (subs-annot-long,11.19)};
\addplot [draw=none,fill=gray!70] coordinates {(subs-annot,9.89) (subs-annot-long,9.78)};
\addplot [draw=none,fill=black!70] coordinates {(subs-annot,9.27) (subs-annot-long,9.30)};
\addplot [draw=none,fill=black!70,postaction={pattern=north west lines,pattern color=gray!30}] coordinates {(subs-annot,8.80) (subs-annot-long,8.63)};
\end{axis}
\end{tikzpicture}%
}

\bigskip
\subfloat[c][Subtitling  \label{fig:scaling_bleu}]{%
\begin{tikzpicture}
\begin{axis}[
        ybar, width=\textwidth, height=0.3\columnwidth,
        bar width=0.6cm, ymin=0, ymax=70,
        axis x line*=bottom, axis y line*=left,
        y axis line style={opacity=0.5},
	ylabel={BLEU (\%)}, ylabel style={align=center, font=\footnotesize},
        yticklabel style={font=\footnotesize},
        xticklabel style={font=\footnotesize},
        enlarge x limits=0.6,
        symbolic x coords={subs-valid,subs-valid-14kh},
        xtick=data,
        xticklabels={\textit{subs-valid},\textit{subs-valid-14kh}},
        nodes near coords,
        every node near coord/.append style={font=\tiny},
        nodes near coords align={vertical},
]
\addplot [draw=gray!20,fill=gray!20,pattern={north east lines},pattern color=gray] coordinates {(subs-valid,32.01) (subs-valid-14kh,34.06)};
\addplot [draw=none,fill=gray!20] coordinates {(subs-valid,37.59) (subs-valid-14kh,42.28)};
\addplot [draw=none,fill=gray!70] coordinates {(subs-valid,48.70) (subs-valid-14kh,55.17)};
\addplot [draw=none,fill=black!70] coordinates {(subs-valid,51.10) (subs-valid-14kh,59.40)};
\addplot [draw=none,fill=black!70,postaction={pattern=north west lines,pattern color=gray!30}] coordinates {(subs-valid,53.87) (subs-valid-14kh,61.76)};
\end{axis}
\end{tikzpicture}%
}
\center{\ref{named_legend_2}}
\caption{Scaling experiments for the cascaded model with dual encoder features. On the horizontal axis, increasing amounts of subtitled training data are used. Figure (a) shows WERs ($\downarrow$) of the verbatim ASR decoder outputs with respect to the reference verbatim transcription. Figure (b) shows BLEU scores ($\uparrow$) of the subtitle decoder outputs with respect to the reference subtitles. All pairwise comparisons between the results are statistically significant.}
\label{fig:scaling}
\end{figure*}

\noindent The cascaded model with dual encoder features from Figure~\ref{fig:cascadedual}, which performed best in Table~\ref{tab:comb1}, is used to perform a scaling experiment on the large subtitle dataset of 14k hours \textit{subs-14kh}, with enriched combined utterance transcriptions. We generate 4 different subsets of the dataset with an increasing amount of subtitle data. The verbatim data is in every experiment a 3-fold speed perturbed version of the verbatim CGN dataset. Figure~\ref{fig:scaling} shows the verbatim results in terms of WER on the short-form \textit{cgn-dev} and \textit{subs-annot} and the long-form \textit{cgn-dev-long} and \textit{subs-annot-long}, and the subtitling results in terms of BLEU score on \textit{subs-valid} and \textit{subs-valid-14kh}. The \textit{XL-model} in Figure~\ref{fig:scaling} is a larger version of the model as described in Section~\ref{sec:model}.

All figures indicate that the inclusion of more subtitled data leads to an improved performance, showing that the proposed method is scalable. For the verbatim speech recognition experiments, we note unseen scores with up to 50\% relative WER reduction compared to the baseline ASR model on the test sets, without adding any additional verbatim data to the verbatim ASR decoder. For large-scale data, the larger \textit{XL-model} is better able to learn all variations in the data.
As expected, the achieved WER reductions do not scale linearly with the amount of weakly supervised data (note that this is still a relatively small model). However, models with larger capacity and training time would probably reach even higher returns.
Finally, Figure~\ref{fig:scaling_bleu} shows the impressive subtitle generation capabilities of the proposed model. We remark that the \textit{subs-valid-14kh} evaluation set is a very difficult set with random held-out samples from the entire 14000 hour corpus, containing a very broad spectrum of speech types and dialects. The cascaded model trained on \textit{subs-14kh} exhibits very high BLEU scores, which can be interpreted as strong translations.

\subsubsection{Comparison to Whisper}
\noindent Finally, we compare our proposed model to the state-of-the-art multilingual speech recognition model Whisper \cite{whisper}. Whisper is an encoder-decoder multi-task speech model that is able to transcribe speech as well as translate speech to English text. It is trained on 680k hours (or more, for \textit{Whisper-large-v3}) of weakly labelled speech data. Since Whisper is trained on long-form audio (30 second chunks), for a fair comparison we evaluate on the long-form test sets. Table~\ref{tab:asrfinal} shows the WERs on \textit{cgn-dev-long} and \textit{subs-annot-long}, comparing our proposed method to Whisper. The first row corresponds to the baseline encoder-decoder ASR model trained on long-form supervised data from the CGN dataset only (second row of Table~\ref{tab:comb1}). For our proposed methods, we use the cascaded model with dual encoder features that was trained on the combination of CGN and 14k hours of subtitle data from Section~\ref{sec:scaling}, and include both the base model and the XL model. These models are compared to the Whisper model without finetuning (\textit{Whisper-large-v3})\footnote{\url{https://huggingface.co/openai/whisper-large-v3}}, and to a finetuned version of the Whisper model that was finetuned on the CGN dataset (i.e. a finetuned version of \textit{Whisper-large-v2})\footnote{\url{https://huggingface.co/kul-speech-lab/whisper_large_CGN}}.

\begin{table}
\begin{center}
\caption{WERs (\%) of verbatim speech recognition experiments ($\downarrow$) with comparison to Whisper. The models are evaluated on \textit{cgn-dev-long} and \textit{subs-annot-long}. The second column denotes the number of parameters in every model. \\}
\label{tab:asrfinal}
\footnotesize
\begin{tabular}{@{\extracolsep{2pt}} l | c | c  c @{}}
\toprule
\textbf{Model} & \# Params. & \textit{cgn-dev-long}~ & \textit{subs-annot-long}~ \\
\midrule
E2E ASR & 50M & 10.61~ & 14.17~ \\
\hline
Proposed Model & 70M & 6.93~ & 9.30~ \\
Proposed Model (XL) & 180M & \textbf{6.49}~ & \textbf{8.63}~ \\
\hline
Whisper large & 1550M & 11.54~ & 13.76~ \\
Whisper large finetuned & 1550M & 7.83~ & 10.64~ \\
\bottomrule 
\end{tabular}
\end{center}
\end{table}

The results in Table~\ref{tab:asrfinal} show the benefits of the proposed approach and the synergistic effect of using subtitles in the same language on the resulting verbatim speech recognition performance. While Whisper is trained on a huge amount of speech data, it lacks the distinction between verbatim transcripts and subtitle transcripts, and is outperformed by our method in this verbatim speech recognition task. Even after finetuning the high-capacity Whisper model on verbatim Dutch data, our proposed model reaches better WERs, as a result of the subtitle data. If the Whisper model were also to be finetuned on the entire subtitle dataset (for which we do not have the required resources), it's output would probably differ strongly from verbatim transcripts, leading to high WERs on this task. 

Moreover, our proposed model is very efficient in terms of parameters and compute resources compared to the large Whisper model, using only about 10 percent of it's parameters. In experiments not reported here, we found that smaller Whisper models perform much worse.

In addition to producing verbatim transcripts, our methods are able to generate subtitles at the same time. We have analysed whether cascading LLMs to ASR outputs leads to subtitles of similar quality as our proposed methods, which are able to condition on the audio as well. We found that the produced subtitles with the joint model outperforms a general ASR+LLM pipeline. Detailed results can be found in \ref{sec:app-llm}.

\section{Discussion}   \label{sec:discussion}

\noindent The experiments in Section~\ref{sec:exp} have shown that the proposed methods are able to leverage subtitle data to improve ASR models, through disambiguation between the two domains. In most cases, the cascaded models outperform the other approaches. As Table~\ref{tab:comb1} illustrates, the cascaded model with dual encoder features generally yields the best ASR performance. There seems to be no harm in combining a large body of subtitle data with a small set of verbatim data. Moreover, the subtitles bring about a positive learning effect which even enhances the verbatim transcription, as the encoder is improved. The proposed approach offers up to 20 percent relative improvement on in-domain test data using only 720 hours of subtitled data, which can be increased to even 50 percent when using more data. On top of that, the domain mismatch between the speech in the ASR dataset and the (often more spontaneous) speech on TV is relieved.

Furthermore, the proposed models are able to generate subtitles that contain many translational effects with regards to written text (Standard Dutch). We notice that even if the verbatim transcription on an out-of-domain sample is far from perfect, often due to a strong local dialect, the subtitle decoder is in most cases able to grasp the content fairly well, as it can be trained more easily on lots of data. Additionally, the experiments with large-scale data have shown very good results on a broad domain, with a performance that is suitable for automatic subtitling. Reaching optimal performance on all local variations probably requires more balancing of the dataset with respect to local dialects and overrepresented speakers when the data, and therefore also the required computing resources, become very large. 

Finally, as the proposed models generate both a verbatim transcription and a subtitle, they can be used for multiple applications. Depending on the use case, either the verbatim transcript (e.g. for speaker analysis, note taking) or the subtitle (e.g. for multimedia) can be useful. For NLP applications, the subtitle outputs should also work better as they are closer to standard written text than verbatim ASR outputs.

\section{Conclusion and Future Work}   \label{sec:conclusion}
\noindent We proposed several architectures to improve automatic speech recognition with weakly supervised data in the form of TV subtitles. The proposed models are able to generate both a verbatim transcription and a subtitle for a spoken utterance. A cascaded encoder approach with separate decoders that attend to both encoder outputs shows a strong learning effect from the subtitle data to the verbatim branch. Evaluation has shown improvements on both verbatim ASR and on automatic subtitling of broadcast TV shows. We carried out scaling experiments on a large subtitle database to prove the scalability of the methods. This work has resulted in a new state-of-the-art ASR model for Belgian Dutch, a language with only a limited amount of manually annotated verbatim training data. Moreover, the proposed approach provides a general architecture that can cope with approximate, weakly supervised transcripts.

For future work, we are interested in investigating verbatim pseudo-labeling approaches to generate parallel data to learn even stronger connections between the different modalities. Furthermore, as there are many datasets containing subtitle texts (e.g. from open-source movie subtitles), a language model can be built with these sources to improve the final transcriptions. It might also be interesting to incorporate timestamp prediction in the current framework to predict subtitle timings on the fly. Moreover, a multilingual dual model can be trained by leveraging subtitles and ASR datasets from many languages. Finally, future work could evaluate the impact of weakly supervised transcripts for ASR models that consist of a combination of a speech encoder and an LLM decoder.

\section*{Acknowledgement}
\noindent This research was supported by Research Foundation Flanders (FWO) under grant S004923N of the SBO programme and by the Flanders AI Impulse Programme - FAIR2.0. The computing resources and services used in this work were provided by the VSC (Flemish Supercomputer Center), funded by the Research Foundation Flanders (FWO) and the Flemish Government. We also thank VRT for the data resources.

\bibliographystyle{elsarticle-num} 
\renewcommand*{\bibfont}{\normalfont\small}
\bibliography{refs.bib}

\begin{thebibliography}{10}
\expandafter\ifx\csname url\endcsname\relax
  \def\url#1{\texttt{#1}}\fi
\expandafter\ifx\csname urlprefix\endcsname\relax\def\urlprefix{URL }\fi
\expandafter\ifx\csname href\endcsname\relax
  \def\href#1#2{#2} \def\path#1{#1}\fi

\bibitem{vaswani}
A.~Vaswani, et~al., \href{https://proceedings.neurips.cc/paper_files/paper/2017/file/3f5ee243547dee91fbd053c1c4a845aa-Paper.pdf}{Attention is all you need}, in: Proc. Conf. on Neural Information Processing Systems (NeurIPS), 2017, pp. 5998--6008.
\newline\urlprefix\url{https://proceedings.neurips.cc/paper_files/paper/2017/file/3f5ee243547dee91fbd053c1c4a845aa-Paper.pdf}

\bibitem{conformer}
A.~Gulati, et~al., Conformer: Convolution-augmented transformer for speech recognition, in: Proc. Interspeech, 2020, pp. 5036--5040.
\newblock \href {https://doi.org/10.21437/Interspeech.2020-3015} {\path{doi:10.21437/Interspeech.2020-3015}}.

\bibitem{bigssl}
Y.~Zhang, et~al., {BigSSL}: Exploring the frontier of large-scale semi-supervised learning for automatic speech recognition, IEEE Journal of Selected Topics in Signal Processing (JSTSP) 16~(6) (2022) 1519--1532.
\newblock \href {https://doi.org/10.1109/JSTSP.2022.3182537} {\path{doi:10.1109/JSTSP.2022.3182537}}.

\bibitem{zhang2022pushing}
Y.~Zhang, et~al., Pushing the limits of semi-supervised learning for automatic speech recognition, in: Proc. Conf. on Neural Information Processing Systems (NeurIPS): SAS Workshop, 2022.

\bibitem{xlsr}
A.~Babu, et~al., {XLS-R}: Self-supervised cross-lingual speech representation learning at scale, in: Proc. Interspeech, 2022, pp. 2278--2282.
\newblock \href {https://doi.org/10.21437/Interspeech.2022-143} {\path{doi:10.21437/Interspeech.2022-143}}.

\bibitem{feng2024}
S.~Feng, B.~M. Halpern, O.~Kudina, O.~Scharenborg, \href{https://www.sciencedirect.com/science/article/pii/S0885230823000864}{Towards inclusive automatic speech recognition}, Computer, Speech and Language 84 (2024) 101567.
\newblock \href {https://doi.org/https://doi.org/10.1016/j.csl.2023.101567} {\path{doi:https://doi.org/10.1016/j.csl.2023.101567}}.
\newline\urlprefix\url{https://www.sciencedirect.com/science/article/pii/S0885230823000864}

\bibitem{chen23l_interspeech}
W.~Chen, X.~Chang, Y.~Peng, Z.~Ni, S.~Maiti, S.~Watanabe, Reducing barriers to self-supervised learning: {HuBERT} pre-training with academic compute, in: Proc. Interspeech, 2023, pp. 4404--4408.
\newblock \href {https://doi.org/10.21437/Interspeech.2023-1176} {\path{doi:10.21437/Interspeech.2023-1176}}.

\bibitem{Chung2021w2vBERTCC}
Y.-A. Chung, et~al., {w2v-BERT}: Combining contrastive learning and masked language modeling for self-supervised speech pre-training, in: Proc. IEEE Automatic Speech Recognition and Understanding Workshop (ASRU), 2021, pp. 244--250.
\newblock \href {https://doi.org/10.1109/ASRU51503.2021.9688253} {\path{doi:10.1109/ASRU51503.2021.9688253}}.

\bibitem{pmlr-v162-chiu22a}
C.-C. Chiu, J.~Qin, Y.~Zhang, J.~Yu, Y.~Wu, Self-supervised learning with random-projection quantizer for speech recognition, in: Proc. Int. Conf. on Machine Learning (ICML), 2022, pp. 3915--3924.

\bibitem{ssl_review}
A.~Mohamed, et~al., Self-supervised speech representation learning: A review, IEEE Journal of Selected Topics in Signal Processing (JSTSP) 16~(6) (2022) 1179--1210.
\newblock \href {https://doi.org/10.1109/JSTSP.2022.3207050} {\path{doi:10.1109/JSTSP.2022.3207050}}.

\bibitem{baevski2020wav2vec}
A.~Baevski, H.~Zhou, A.~Mohamed, M.~Auli, Wav2vec 2.0: A framework for self-supervised learning of speech representations, in: Proc. Conf. on Neural Information Processing Systems (NeurIPS), 2020, pp. 12449--12460.

\bibitem{hsu2021hubert}
W.-N. Hsu, B.~Bolte, Y.-H.~H. Tsai, K.~Lakhotia, R.~Salakhutdinov, A.~Mohamed, {HuBERT}: Self-supervised speech representation learning by masked prediction of hidden units, IEEE/ACM Trans. on Audio, Speech, and Language Processing 29 (2021) 3451--3460.
\newblock \href {https://doi.org/10.1109/TASLP.2021.3122291} {\path{doi:10.1109/TASLP.2021.3122291}}.

\bibitem{wavlm}
S.~Chen, et~al., {WavLM}: Large-scale self-supervised pre-training for full stack speech processing, IEEE Journal of Selected Topics in Signal Processing (JSTSP) 16~(6) (2022) 1505--1518.
\newblock \href {https://doi.org/10.1109/JSTSP.2022.3188113} {\path{doi:10.1109/JSTSP.2022.3188113}}.

\bibitem{conneau21_interspeech}
A.~Conneau, A.~Baevski, R.~Collobert, A.~Mohamed, M.~Auli, Unsupervised cross-lingual representation learning for speech recognition, in: Proc. Interspeech, 2021, pp. 2426--2430.
\newblock \href {https://doi.org/10.21437/Interspeech.2021-329} {\path{doi:10.21437/Interspeech.2021-329}}.

\bibitem{mhubert}
A.~Lee, et~al., Textless speech-to-speech translation on real data, in: Proc. Conf. North {A}merican Chapter of the Association for Computational Linguistics (NAACL): Human Language Technologies, 2022, pp. 860--872.

\bibitem{wavlablm}
W.~Chen, et~al., Joint prediction and denoising for large-scale multilingual self-supervised learning, in: Proc. IEEE Automatic Speech Recognition and Understanding Workshop (ASRU), 2023.

\bibitem{superb}
S.~Yang, et~al., {SUPERB}: Speech processing universal performance benchmark, in: Proc. Interspeech, 2021, pp. 1194--1198.

\bibitem{mlsuperb}
J.~Shi, et~al., {ML-SUPERB}: Multilingual speech universal performance benchmark, in: Proc. Interspeech, 2023, pp. 884--888.
\newblock \href {https://doi.org/10.21437/Interspeech.2023-1316} {\path{doi:10.21437/Interspeech.2023-1316}}.

\bibitem{hsu21_interspeech}
W.-N. Hsu, et~al., Robust wav2vec 2.0: Analyzing domain shift in self-supervised pre-training, in: Proc. Interspeech, 2021, pp. 721--725.
\newblock \href {https://doi.org/10.21437/Interspeech.2021-236} {\path{doi:10.21437/Interspeech.2021-236}}.

\bibitem{whisper}
A.~Radford, J.~W. Kim, T.~Xu, G.~Brockman, C.~Mcleavey, I.~Sutskever, Robust speech recognition via large-scale weak supervision, in: Proc. Int. Conf. on Machine Learning (ICML), 2023, pp. 28492--28518.

\bibitem{speechstew}
W.~Chan, D.~S. Park, C.~A. Lee, Y.~Zhang, Q.~V. Le, M.~Norouzi, {SpeechStew}: Simply mix all available speech recognition data to train one large neural network, in: Workshop on Machine Learning in Speech and Language Processing (MLSLP), 2021.

\bibitem{owsm}
Y.~Peng, et~al., Reproducing {Whisper}-style training using an open-source toolkit and publicly available data, in: Proc. IEEE Automatic Speech Recognition and Understanding Workshop (ASRU), 2023.

\bibitem{likhomanenko21_interspeech}
T.~Likhomanenko, et~al., Rethinking evaluation in {ASR}: Are our models robust enough?, in: Proc. Interspeech, 2021, pp. 311--315.
\newblock \href {https://doi.org/10.21437/Interspeech.2021-1758} {\path{doi:10.21437/Interspeech.2021-1758}}.

\bibitem{cintas2014audiovisual}
J.~D. Cintas, A.~Remael, Audiovisual translation: Subtitling, Routledge, 2014.

\bibitem{MustCinema}
A.~Karakanta, M.~Negri, M.~Turchi, {M}u{ST}-{C}inema: a speech-to-subtitles corpus, in: Proc. Int. Conf. on Language Resources and Evaluation (LREC), 2020, pp. 3727--3734.

\bibitem{librispeech}
V.~Panayotov, G.~Chen, D.~Povey, S.~Khudanpur, {LibriSpeech}: An {ASR} corpus based on public domain audio books, in: Proc. IEEE Int. Conf. on Acoustics, Speech and Signal Processing (ICASSP), 2015, pp. 5206--5210.

\bibitem{roy09b_interspeech}
B.~C. Roy, D.~Roy, Fast transcription of unstructured audio recordings, in: Proc. Interspeech, 2009, pp. 1647--1650.
\newblock \href {https://doi.org/10.21437/Interspeech.2009-500} {\path{doi:10.21437/Interspeech.2009-500}}.

\bibitem{reazonspeech}
Y.~Yin, D.~Mori, S.~Fujimoto, {ReazonSpeech}: A free and massive corpus for {J}apanese {ASR}, in: Proc. 29th Annual Meeting of the Association for Natural Language Processing, 2023, pp. 1134--1139.

\bibitem{SubtitleModel}
J.~Poncelet, H.~Van~hamme, Learning to jointly transcribe and subtitle for end-to-end spontaneous speech recognition, in: Proc. IEEE Spoken Language Technology Workshop (SLT), 2023, pp. 182--189.
\newblock \href {https://doi.org/10.1109/SLT54892.2023.10022420} {\path{doi:10.1109/SLT54892.2023.10022420}}.

\bibitem{ihori23_interspeech}
M.~Ihori, H.~Sato, T.~Tanaka, R.~Masumura, S.~Mizuno, N.~Hojo, Transcribing speech as spoken and written dual text using an autoregressive model, in: Proc. Interspeech, 2023, pp. 461--465.
\newblock \href {https://doi.org/10.21437/Interspeech.2023-1655} {\path{doi:10.21437/Interspeech.2023-1655}}.

\bibitem{kanda20b_interspeech}
N.~Kanda, Y.~Gaur, X.~Wang, Z.~Meng, T.~Yoshioka, Serialized output training for end-to-end overlapped speech recognition, in: Proc. Interspeech, 2020, pp. 2797--2801.
\newblock \href {https://doi.org/10.21437/Interspeech.2020-999} {\path{doi:10.21437/Interspeech.2020-999}}.

\bibitem{xu-etal-2022-joint}
J.~Xu, F.~Buet, J.~Crego, E.~Bertin-Lem{\'e}e, F.~Yvon, Joint generation of captions and subtitles with dual decoding, in: Proc. Int. Conf. on Spoken Language Translation (IWSLT), ACL, 2022, pp. 74--82.
\newblock \href {https://doi.org/10.18653/v1/2022.iwslt-1.7} {\path{doi:10.18653/v1/2022.iwslt-1.7}}.

\bibitem{lamel_2002}
L.~Lamel, J.-L. Gauvain, G.~Adda, Lightly supervised and unsupervised acoustic model training, Computer, Speech and Language 16~(1) (2002) 115--129.
\newblock \href {https://doi.org/https://doi.org/10.1006/csla.2001.0186} {\path{doi:https://doi.org/10.1006/csla.2001.0186}}.

\bibitem{lanchantin16_interspeech}
P.~Lanchantin, et~al., Selection of multi-genre broadcast data for the training of automatic speech recognition systems, in: Proc. Interspeech, 2016, pp. 3057--3061.
\newblock \href {https://doi.org/10.21437/Interspeech.2016-462} {\path{doi:10.21437/Interspeech.2016-462}}.

\bibitem{BangIEICE}
J.-U. Bang, M.-Y. Choi, S.-H. Kim, O.-W. Kwon, Automatic construction of a large-scale speech recognition database using multi-genre broadcast data with inaccurate subtitle timestamps, IEICE Trans. on Information and Systems E103.D~(2) (2020) 406--415.
\newblock \href {https://doi.org/10.1587/transinf.2019EDP7234} {\path{doi:10.1587/transinf.2019EDP7234}}.

\bibitem{Ando2021ConstructionOA}
S.~Ando, H.~Fujihara, Construction of a large-scale {Japanese} {ASR} corpus on {TV} recordings, in: Proc. IEEE Int. Conf. on Acoustics, Speech and Signal Processing (ICASSP), 2021, pp. 6948--6952.

\bibitem{bell15_ASRU}
P.~Bell, S.~Renals, A system for automatic alignment of broadcast media captions using weighted finite-state transducers, in: Proc. IEEE Automatic Speech Recognition and Understanding Workshop (ASRU), 2015, pp. 675--680.
\newblock \href {https://doi.org/10.1109/ASRU.2015.7404861} {\path{doi:10.1109/ASRU.2015.7404861}}.

\bibitem{Saz2018}
O.~Saz, et~al., Lightly supervised alignment of subtitles on multi-genre broadcasts, in: Multimedia Tools and Applications, Vol.~77, 2018, pp. 30533--30550.
\newblock \href {https://doi.org/10.1007/s11042-018-6050-1} {\path{doi:10.1007/s11042-018-6050-1}}.

\bibitem{JHU_kaldi}
V.~Manohar, D.~Povey, S.~Khudanpur, {JHU} {K}aldi system for {A}rabic {MGB-3} {ASR} challenge using diarization, audio-transcript alignment and transfer learning, in: Proc. IEEE Automatic Speech Recognition and Understanding Workshop (ASRU), 2017, pp. 346--352.
\newblock \href {https://doi.org/10.1109/ASRU.2017.8268956} {\path{doi:10.1109/ASRU.2017.8268956}}.

\bibitem{vishwa2015}
V.~Gupta, P.~Deléglise, G.~Boulianne, Y.~Estève, S.~Meignier, A.~Rousseau, {CRIM} and {LIUM} approaches for multi-genre broadcast media transcription, in: Proc. IEEE Automatic Speech Recognition and Understanding Workshop (ASRU), 2015, pp. 681--686.
\newblock \href {https://doi.org/10.1109/ASRU.2015.7404862} {\path{doi:10.1109/ASRU.2015.7404862}}.

\bibitem{Guerreiro21eswa}
N.~M. Guerreiro, R.~Rei, F.~Batista, Towards better subtitles: A multilingual approach for punctuation restoration of speech transcripts, Expert Systems with Applications 186 (2021) 115740.
\newblock \href {https://doi.org/https://doi.org/10.1016/j.eswa.2021.115740} {\path{doi:https://doi.org/10.1016/j.eswa.2021.115740}}.

\bibitem{geislinger-etal-2022-improved}
R.~Geislinger, B.~Milde, C.~Biemann, Improved open source automatic subtitling for lecture videos, in: Proc. Conf. on Natural Language Processing (KONVENS), 2022, pp. 98--103.

\bibitem{Milde2021_1109}
B.~Milde, R.~Geislinger, I.~Lindt, T.~Baumann, Open source automatic lecture subtitling, in: Proc. Conf. on Electronical Speech Signal Processing (ESSV), 2021, pp. 128--135.

\bibitem{liu-etal-2020-adapting}
D.~Liu, J.~Niehues, G.~Spanakis, Adapting end-to-end speech recognition for readable subtitles, in: Proc. Int. Conf. on Spoken Language Translation (IWSLT), ACL, 2020, pp. 247--256.
\newblock \href {https://doi.org/10.18653/v1/2020.iwslt-1.30} {\path{doi:10.18653/v1/2020.iwslt-1.30}}.

\bibitem{MGB1}
P.~Bell, et~al., The {MGB} challenge: Evaluating multi-genre broadcast media recognition, in: Proc. IEEE Automatic Speech Recognition and Understanding Workshop (ASRU), 2015, pp. 687--693.
\newblock \href {https://doi.org/10.1109/ASRU.2015.7404863} {\path{doi:10.1109/ASRU.2015.7404863}}.

\bibitem{MGB2}
A.~Ali, et~al., The {MGB}-2 challenge: {Arabic} multi-dialect broadcast media recognition, in: Proc. IEEE Spoken Language Technology Workshop (SLT), 2016, pp. 279--284.
\newblock \href {https://doi.org/10.1109/SLT.2016.7846277} {\path{doi:10.1109/SLT.2016.7846277}}.

\bibitem{iberspeech}
E.~Lleida, et~al., \href{https://www.mdpi.com/2076-3417/9/24/5412}{Albayzin 2018 evaluation: The {IberSpeech-RTVE} challenge on speech technologies for {Spanish} broadcast media}, Applied Sciences 9~(24) (2019).
\newblock \href {https://doi.org/10.3390/app9245412} {\path{doi:10.3390/app9245412}}.
\newline\urlprefix\url{https://www.mdpi.com/2076-3417/9/24/5412}

\bibitem{che2017}
X.~Che, S.~Luo, H.~Yang, C.~Meinel, Automatic lecture subtitle generation and how it helps, in: Proc. IEEE Int. Conf. on Advanced Learning Technologies (ICALT), 2017, pp. 34--38.
\newblock \href {https://doi.org/10.1109/ICALT.2017.11} {\path{doi:10.1109/ICALT.2017.11}}.

\bibitem{papi-etal-2023-direct-speech}
S.~Papi, M.~Gaido, A.~Karakanta, M.~Cettolo, M.~Negri, M.~Turchi, \href{https://aclanthology.org/2023.tacl-1.77}{Direct speech translation for automatic subtitling}, Trans. of the Assoc. for Computational Linguistics 11 (2023) 1355--1376.
\newblock \href {https://doi.org/10.1162/tacl_a_00607} {\path{doi:10.1162/tacl_a_00607}}.
\newline\urlprefix\url{https://aclanthology.org/2023.tacl-1.77}

\bibitem{kurzinger2020}
L.~K{\"u}rzinger, D.~Winkelbauer, L.~Li, T.~Watzel, G.~Rigoll, {CTC}-segmentation of large corpora for {German} end-to-end speech recognition, in: Int. Conf. on Speech and Computer (SPECOM), 2020, pp. 267--278.

\bibitem{ihori-etal-2020-parallel}
M.~Ihori, A.~Takashima, R.~Masumura, Parallel corpus for {J}apanese spoken-to-written style conversion, in: Proc. Int. Conf. on Language Resources and Evaluation (LREC), 2020, pp. 6346--6353.

\bibitem{liao23}
J.~Liao, et~al., Improving readability for automatic speech recognition transcription, ACM Trans. on Asian and Low-Resource Language Information Processing 22~(5) (2023).
\newblock \href {https://doi.org/10.1145/3557894} {\path{doi:10.1145/3557894}}.

\bibitem{nozaki22_interspeech}
J.~Nozaki, T.~Kawahara, K.~Ishizuka, T.~Hashimoto, End-to-end speech-to-punctuated-text recognition, in: Proc. Interspeech, 2022, pp. 1811--1815.
\newblock \href {https://doi.org/10.21437/Interspeech.2022-5} {\path{doi:10.21437/Interspeech.2022-5}}.

\bibitem{futami23_icassp}
H.~Futami, et~al., Streaming joint speech recognition and disfluency detection, in: Proc. IEEE Int. Conf. on Acoustics, Speech and Signal Processing (ICASSP), 2023.
\newblock \href {https://doi.org/10.1109/ICASSP49357.2023.10094620} {\path{doi:10.1109/ICASSP49357.2023.10094620}}.

\bibitem{sunkara21_icassp}
M.~Sunkara, C.~Shivade, S.~Bodapati, K.~Kirchhoff, Neural inverse text normalization, in: Proc. IEEE Int. Conf. on Acoustics, Speech and Signal Processing (ICASSP), 2021, pp. 7573--7577.
\newblock \href {https://doi.org/10.1109/ICASSP39728.2021.9414912} {\path{doi:10.1109/ICASSP39728.2021.9414912}}.

\bibitem{wang-etal-2022-adaptive}
Z.~Wang, Y.~Wang, S.~Wang, W.~Che, Adaptive unsupervised self-training for disfluency detection, in: Proc. Int. Conf. on Computational Linguistics (COLING), ICCL, 2022, pp. 7209--7218.

\bibitem{guo19_icassp}
J.~Guo, T.~N. Sainath, R.~J. Weiss, A spelling correction model for end-to-end speech recognition, in: Proc. IEEE Int. Conf. on Acoustics, Speech and Signal Processing (ICASSP), 2019, pp. 5651--5655.
\newblock \href {https://doi.org/10.1109/ICASSP.2019.8683745} {\path{doi:10.1109/ICASSP.2019.8683745}}.

\bibitem{li2017}
S.~Li, X.~Lu, S.~Sakai, M.~Mimura, T.~Kawahara, Semi-supervised ensemble {DNN} acoustic model training, in: Proc. IEEE Int. Conf. on Acoustics, Speech and Signal Processing (ICASSP), 2017, pp. 5270--5274.
\newblock \href {https://doi.org/10.1109/ICASSP.2017.7953162} {\path{doi:10.1109/ICASSP.2017.7953162}}.

\bibitem{li2019}
B.~Li, T.~N. Sainath, R.~Pang, Z.~Wu, Semi-supervised training for end-to-end models via weak distillation, in: Proc. IEEE Int. Conf. on Acoustics, Speech and Signal Processing (ICASSP), 2019, pp. 2837--2841.
\newblock \href {https://doi.org/10.1109/ICASSP.2019.8682172} {\path{doi:10.1109/ICASSP.2019.8682172}}.

\bibitem{pratap2022_neurips}
V.~Pratap, A.~Hannun, G.~Synnaeve, R.~Collobert, \href{https://proceedings.neurips.cc/paper_files/paper/2022/file/57587d8d6a7ede0e5302fc22d0878c53-Paper-Conference.pdf}{Star {T}emporal {C}lassification: Sequence modeling with partially labeled data}, in: Proc. Conf. on Neural Information Processing Systems (NeurIPS), 2022, pp. 13392--13403.
\newline\urlprefix\url{https://proceedings.neurips.cc/paper_files/paper/2022/file/57587d8d6a7ede0e5302fc22d0878c53-Paper-Conference.pdf}

\bibitem{pratap2022_icassp}
V.~Pratap, Q.~Xu, T.~Likhomanenko, G.~Synnaeve, R.~Collobert, Word order does not matter for speech recognition, in: Proc. IEEE Int. Conf. on Acoustics, Speech and Signal Processing (ICASSP), 2022, pp. 7202--7206.
\newblock \href {https://doi.org/10.1109/ICASSP43922.2022.9747805} {\path{doi:10.1109/ICASSP43922.2022.9747805}}.

\bibitem{sing2020_icassp}
K.~Singh, et~al., Training {ASR} models by generation of contextual information, in: Proc. IEEE Int. Conf. on Acoustics, Speech and Signal Processing (ICASSP), 2020, pp. 7864--7868.
\newblock \href {https://doi.org/10.1109/ICASSP40776.2020.9053527} {\path{doi:10.1109/ICASSP40776.2020.9053527}}.

\bibitem{chen21o_interspeech}
G.~Chen, et~al., {GigaSpeech}: An evolving, multi-domain {ASR} corpus with 10,000 hours of transcribed audio, in: Proc. Interspeech, 2021, pp. 3670--3674.
\newblock \href {https://doi.org/10.21437/Interspeech.2021-1965} {\path{doi:10.21437/Interspeech.2021-1965}}.

\bibitem{wenetspeech}
B.~Zhang, et~al., {WenetSpeech}: A 10000+ hours multi-domain {Mandarin} corpus for speech recognition, in: Proc. IEEE Int. Conf. on Acoustics, Speech and Signal Processing (ICASSP), 2022, pp. 6182--6186.
\newblock \href {https://doi.org/10.1109/ICASSP43922.2022.9746682} {\path{doi:10.1109/ICASSP43922.2022.9746682}}.

\bibitem{galvez2021peoples}
D.~Galvez, et~al., \href{https://datasets-benchmarks-proceedings.neurips.cc/paper_files/paper/2021/file/202cb962ac59075b964b07152d234b70-Paper-round1.pdf}{The {People’s} {Speech}: A large-scale diverse {English} speech recognition dataset for commercial usage}, in: Proc. Conf. on Neural Information Processing Systems (NeurIPS): Track on Datasets and Benchmarks, 2021.
\newline\urlprefix\url{https://datasets-benchmarks-proceedings.neurips.cc/paper_files/paper/2021/file/202cb962ac59075b964b07152d234b70-Paper-round1.pdf}

\bibitem{wilken-etal-2022-suber}
P.~Wilken, P.~Georgakopoulou, E.~Matusov, \href{https://aclanthology.org/2022.iwslt-1.1}{{S}ub{ER}: A metric for automatic evaluation of subtitle quality}, in: Proc. Int. Conf. on Spoken Language Translation (IWSLT), ACL, 2022, pp. 1--10.
\newblock \href {https://doi.org/10.18653/v1/2022.iwslt-1.1} {\path{doi:10.18653/v1/2022.iwslt-1.1}}.
\newline\urlprefix\url{https://aclanthology.org/2022.iwslt-1.1}

\bibitem{karakanta-etal-2022-evaluating}
A.~Karakanta, F.~Buet, M.~Cettolo, F.~Yvon, \href{https://aclanthology.org/2022.lrec-1.328}{Evaluating subtitle segmentation for end-to-end generation systems}, in: Proc. Int. Conf. on Language Resources and Evaluation (LREC), ELRA, 2022, pp. 3069--3078.
\newline\urlprefix\url{https://aclanthology.org/2022.lrec-1.328}

\bibitem{ctc}
A.~Graves, S.~Fern\'{a}ndez, F.~Gomez, J.~Schmidhuber, \href{https://doi.org/10.1145/1143844.1143891}{Connectionist {T}emporal {C}lassification: Labelling unsegmented sequence data with recurrent neural networks}, in: Proc. Int. Conf. on Machine Learning (ICML), ACM, 2006, p. 369–376.
\newblock \href {https://doi.org/10.1145/1143844.1143891} {\path{doi:10.1145/1143844.1143891}}.
\newline\urlprefix\url{https://doi.org/10.1145/1143844.1143891}

\bibitem{interctc}
J.~Lee, S.~Watanabe, Intermediate loss regularization for {CTC}-based speech recognition, in: Proc. IEEE Int. Conf. on Acoustics, Speech and Signal Processing (ICASSP), 2021, pp. 6224--6228.
\newblock \href {https://doi.org/10.1109/ICASSP39728.2021.9414594} {\path{doi:10.1109/ICASSP39728.2021.9414594}}.

\bibitem{watanabe2017}
S.~Watanabe, T.~Hori, S.~Kim, J.~R. Hershey, T.~Hayashi, Hybrid {CTC}/{A}ttention architecture for end-to-end speech recognition, IEEE Journal of Selected Topics in Signal Processing (JSTSP) 11~(8) (2017) 1240--1253.
\newblock \href {https://doi.org/10.1109/JSTSP.2017.2763455} {\path{doi:10.1109/JSTSP.2017.2763455}}.

\bibitem{smoothing}
C.~Szegedy, V.~Vanhoucke, S.~Ioffe, J.~Shlens, Z.~Wojna, Rethinking the {I}nception architecture for computer vision, in: Proc. IEEE Conf. on Computer Vision and Pattern Recognition (CVPR), 2016, pp. 2818--2826.
\newblock \href {https://doi.org/10.1109/CVPR.2016.308} {\path{doi:10.1109/CVPR.2016.308}}.

\bibitem{tied2018}
A.~Anastasopoulos, D.~Chiang, Tied multitask learning for neural speech translation, in: Proc. Conf. of the North {A}merican Chapter of the Association for Computational Linguistics (NAACL): Human Language Technologies, Vol.~1, 2018, pp. 82--91.
\newblock \href {https://doi.org/10.18653/v1/N18-1008} {\path{doi:10.18653/v1/N18-1008}}.

\bibitem{fastmd}
H.~Inaguma, S.~Dalmia, B.~Yan, S.~Watanabe, Fast-{MD}: Fast multi-decoder end-to-end speech translation with non-autoregressive hidden intermediates, in: Proc. IEEE Automatic Speech Recognition and Understanding Workshop (ASRU), 2021, pp. 922--929.
\newblock \href {https://doi.org/10.1109/ASRU51503.2021.9687894} {\path{doi:10.1109/ASRU51503.2021.9687894}}.

\bibitem{dalmia-etal-2021-searchable}
S.~Dalmia, B.~Yan, V.~Raunak, F.~Metze, S.~Watanabe, \href{https://aclanthology.org/2021.naacl-main.151}{Searchable hidden intermediates for end-to-end models of decomposable sequence tasks}, in: Proc. Conf. of the North {A}merican Chapter of the Association for Computational Linguistics (NAACL): Human Language Technologies, 2021, pp. 1882--1896.
\newblock \href {https://doi.org/10.18653/v1/2021.naacl-main.151} {\path{doi:10.18653/v1/2021.naacl-main.151}}.
\newline\urlprefix\url{https://aclanthology.org/2021.naacl-main.151}

\bibitem{helcl-etal-2018-cuni}
J.~Helcl, J.~Libovick{\'y}, D.~Vari{\v{s}}, {CUNI} system for the {WMT}18 multimodal translation task, in: Proc. Conf. on Machine Translation (WMT), ACL, 2018, pp. 616--623.
\newblock \href {https://doi.org/10.18653/v1/W18-6441} {\path{doi:10.18653/v1/W18-6441}}.

\bibitem{chuang-etal-2021-investigating}
S.-P. Chuang, Y.-S. Chuang, C.-C. Chang, H.-y. Lee, \href{https://aclanthology.org/2021.findings-acl.92}{Investigating the reordering capability in {CTC}-based non-autoregressive end-to-end speech translation}, in: Findings of the Association for Computational Linguistics (ACL-IJCNLP), ACL, 2021, pp. 1068--1077.
\newblock \href {https://doi.org/10.18653/v1/2021.findings-acl.92} {\path{doi:10.18653/v1/2021.findings-acl.92}}.
\newline\urlprefix\url{https://aclanthology.org/2021.findings-acl.92}

\bibitem{yan-etal-2023-ctc}
B.~Yan, et~al., \href{https://aclanthology.org/2023.eacl-main.119}{{CTC} alignments improve autoregressive translation}, in: Proc. Conf. of the European Chapter of the Association for Computational Linguistics, 2023, pp. 1623--1639.
\newblock \href {https://doi.org/10.18653/v1/2023.eacl-main.119} {\path{doi:10.18653/v1/2023.eacl-main.119}}.
\newline\urlprefix\url{https://aclanthology.org/2023.eacl-main.119}

\bibitem{CGN_Oostdijk}
N.~Oostdijk, The {S}poken {D}utch {C}orpus: Overview and first evaluation, in: Proc. Int. Conf. on Language Resources and Evaluation (LREC), Vol.~2, 2000.

\bibitem{ko15_interspeech}
T.~Ko, V.~Peddinti, D.~Povey, S.~Khudanpur, Audio augmentation for speech recognition, in: Proc. Interspeech, 2015, pp. 3586--3589.
\newblock \href {https://doi.org/10.21437/Interspeech.2015-711} {\path{doi:10.21437/Interspeech.2015-711}}.

\bibitem{ponceletASRU}
J.~Poncelet, H.~Van~hamme, Comparison of self-supervised speech pre-training methods on {F}lemish {D}utch, in: Proc. IEEE Automatic Speech Recognition and Understanding Workshop (ASRU), 2021, pp. 169--176.
\newblock \href {https://doi.org/10.1109/ASRU51503.2021.9688061} {\path{doi:10.1109/ASRU51503.2021.9688061}}.

\bibitem{park19e_interspeech}
D.~S. Park, et~al., {SpecAugment}: A simple data augmentation method for automatic speech recognition, in: Proc. Interspeech, 2019, pp. 2613--2617.
\newblock \href {https://doi.org/10.21437/Interspeech.2019-2680} {\path{doi:10.21437/Interspeech.2019-2680}}.

\bibitem{watanabe18_interspeech}
S.~Watanabe, et~al., {ESPnet}: End-to-end speech processing toolkit, in: Proc. Interspeech, 2018, pp. 2207--2211.
\newblock \href {https://doi.org/10.21437/Interspeech.2018-1456} {\path{doi:10.21437/Interspeech.2018-1456}}.

\bibitem{adam}
D.~P. Kingma, J.~Ba, \href{http://arxiv.org/abs/1412.6980}{Adam: {A} method for stochastic optimization}, in: Proc. Int. Conf. on Learning Representations (ICLR), 2015.
\newline\urlprefix\url{http://arxiv.org/abs/1412.6980}

\bibitem{bleu}
K.~Papineni, S.~Roukos, T.~Ward, W.-J. Zhu, \href{https://doi.org/10.3115/1073083.1073135}{{BLEU}: a method for automatic evaluation of machine translation}, in: Proc. 40th Annual Meeting of the Association for Computational Linguistics, 2002, p. 311–318.
\newblock \href {https://doi.org/10.3115/1073083.1073135} {\path{doi:10.3115/1073083.1073135}}.
\newline\urlprefix\url{https://doi.org/10.3115/1073083.1073135}

\bibitem{sacrebleu}
M.~Post, \href{https://www.aclweb.org/anthology/W18-6319}{A call for clarity in reporting {BLEU} scores}, in: Proc. Conf. on Machine Translation (WMT), ACL, 2018, pp. 186--191.
\newline\urlprefix\url{https://www.aclweb.org/anthology/W18-6319}

\bibitem{mapsswe}
D.~Pallet, W.~Fisher, J.~Fiscus, Tools for the analysis of benchmark speech recognition tests, in: Proc. IEEE Int. Conf. on Acoustics, Speech and Signal Processing (ICASSP), 1990, pp. 97--100.
\newblock \href {https://doi.org/10.1109/ICASSP.1990.115546} {\path{doi:10.1109/ICASSP.1990.115546}}.

\bibitem{koehn-2004-statistical}
P.~Koehn, \href{https://aclanthology.org/W04-3250}{Statistical significance tests for machine translation evaluation}, in: Proc. Conf. on Empirical Methods in Natural Language Processing, ACL, 2004, pp. 388--395.
\newline\urlprefix\url{https://aclanthology.org/W04-3250}

\bibitem{vandekerckhove2009_subtitling}
R.~Vandekerckhove, A.~De~Houwer, A.~Remael, \href{https://www.jbe-platform.com/content/journals/10.1075/prag.19.4.05van}{Between language policy and linguistic reality: Intralingual subtitling on {F}lemish television}, Pragmatics 19~(4) (2009) 609--628.
\newblock \href {https://doi.org/https://doi.org/10.1075/prag.19.4.05van} {\path{doi:https://doi.org/10.1075/prag.19.4.05van}}.
\newline\urlprefix\url{https://www.jbe-platform.com/content/journals/10.1075/prag.19.4.05van}

\bibitem{remael2008}
A.~Remael, A.~De~Houwer, R.~Vandekerckhove, \href{https://repository.uantwerpen.be/desktop/irua}{Intralingual open subtitling in {F}landers: audiovisual translation, linguistic variation and audience needs}, Journal of Specialised Translation (JoSTrans) 10 (2008) 76--105.
\newline\urlprefix\url{https://repository.uantwerpen.be/desktop/irua}

\bibitem{prieels2018}
L.~Prieels, G.~De~Sutter, \href{https://www.aup-online.com/content/journals/10.5117/TET2018.2.PRIE}{A mixed-method approach to the use of {C}olloquial {B}elgian {D}utch in intralingual subtitling on {F}lemish television}, Taal en Tongval 70~(2) (2018) 211--256.
\newblock \href {https://doi.org/https://doi.org/10.5117/TET2018.2.PRIE} {\path{doi:https://doi.org/10.5117/TET2018.2.PRIE}}.
\newline\urlprefix\url{https://www.aup-online.com/content/journals/10.5117/TET2018.2.PRIE}

\bibitem{vandekerckhove2009_dialect}
R.~Vandekerckhove, \href{https://doi.org/10.1515/IJSL.2009.017}{Dialect loss and dialect vitality in {F}landers}, Int. Journal of the Sociology of Language 2009~(196-197) (2009) 73--97.
\newblock \href {https://doi.org/doi:10.1515/IJSL.2009.017} {\path{doi:doi:10.1515/IJSL.2009.017}}.
\newline\urlprefix\url{https://doi.org/10.1515/IJSL.2009.017}

\bibitem{vandekerckhove2007}
R.~Vandekerckhove, \href{https://doi.org/10.1515/9783110925463.189}{'{T}ussentaal' as a source of change from below in {B}elgian {D}utch. a case study of substandardization processes in the chat language of {F}lemish teenagers}, in: Germanic Language Histories 'from Below' (1700-2000), De Gruyter, 2007, pp. 189--204.
\newblock \href {https://doi.org/doi:10.1515/9783110925463.189} {\path{doi:doi:10.1515/9783110925463.189}}.
\newline\urlprefix\url{https://doi.org/10.1515/9783110925463.189}

\bibitem{shafey19_interspeech}
L.~E. Shafey, H.~Soltau, I.~Shafran, Joint speech recognition and speaker diarization via sequence transduction, in: Proc. Interspeech, 2019, pp. 396--400.
\newblock \href {https://doi.org/10.21437/Interspeech.2019-1943} {\path{doi:10.21437/Interspeech.2019-1943}}.

\bibitem{kanda22_interspeech}
N.~Kanda, et~al., Streaming multi-talker {ASR} with token-level serialized output training, in: Proc. Interspeech, 2022, pp. 3774--3778.
\newblock \href {https://doi.org/10.21437/Interspeech.2022-7} {\path{doi:10.21437/Interspeech.2022-7}}.

\bibitem{liu2024recordingeyesechoingears}
J.~Liu, C.~Deng, Q.~Zhang, Q.~Chen, H.~Yu, W.~Wang, \href{https://arxiv.org/abs/2408.09688}{Recording for eyes, not echoing to ears: Contextualized spoken-to-written conversion of {ASR} transcripts} (2024).
\newblock \href {http://arxiv.org/abs/2408.09688} {\path{arXiv:2408.09688}}.
\newline\urlprefix\url{https://arxiv.org/abs/2408.09688}

\bibitem{dubey2024llama3herdmodels}
A.~Dubey, et~al., \href{https://arxiv.org/abs/2407.21783}{The {Llama} 3 herd of models} (2024).
\newblock \href {http://arxiv.org/abs/2407.21783} {\path{arXiv:2407.21783}}.
\newline\urlprefix\url{https://arxiv.org/abs/2407.21783}

\bibitem{jiang2023mistral7b}
A.~Q. Jiang, et~al., \href{https://arxiv.org/abs/2310.06825}{Mistral 7b} (2023).
\newblock \href {http://arxiv.org/abs/2310.06825} {\path{arXiv:2310.06825}}.
\newline\urlprefix\url{https://arxiv.org/abs/2310.06825}

\end{thebibliography}

\newpage
\appendix
\section{Dual Outputs: Verbatim and Subtitle Annotations} 

\subsection{Examples of differences between verbatim and subtitle transcripts} \label{sec:app-ex}
\vspace{0.2cm}
\noindent Section~\ref{sec:translation} has explained the general differences between Dutch spoken and written language as well as the differences between verbatim and subtitle annotations. To visualise the dissimilarity and their consequences in practice, Table~\ref{tab:str} shows several examples for speech samples from the \textit{subs-valid-14kh} test set. The examples depict the reference subtitle and a manual verbatim transcription, as well as the predictions from the cascaded model with dual encoder features from Table~\ref{tab:comb1} (using either the subtitle decoder output or the verbatim ASR decoder output respectively). Notice that our model produces both outputs at the same time.

\newcolumntype{C}{>{\centering\arraybackslash}X}
\newcolumntype{L}{>{\raggedright\arraybackslash}X}
\begin{table*}[h]
    \centering
    \footnotesize
    \begin{tabularx}{\columnwidth}{p{1.2cm} L}
        \toprule
        \multirow{2}{=}[-9pt]{\textbf{REF} (verbatim)} & Wacht wacht wacht momentje. Zijt de${<}{*}d{>}$ zeker dat ze d'r niet is? Misschien moet ge eerst efkes checken. \\
        & \textit{Wait wait wait a moment. Are you sure that she's not there? Maybe you just have to check first.} \\
        \midrule[0.01pt]
        \multirow{2}{=}[-9pt]{\textbf{REF} (subtitle)} & Wacht, wacht. Ben je zeker dat ze er niet is? Misschien moet je eerst checken. \\
        & \textit{Wait, wait. Are you sure that she is not there? Maybe you have to check first.} \\
        \midrule[0.01pt]
        \multirow{2}{=}[-9pt]{\textbf{HYP} (verbatim)} & Wacht wacht wacht. Zijt de${<}{*}d{>}$ zeker dat ze d'r niet is? Misschien moet eerst effekes checken? \\
        & \textit{Wait wait wait. Are you sure that she's not there? Maybe you just have to check first?}\\
        \midrule[0.01pt]
        \multirow{2}{=}{\textbf{HYP} (subtitle)} & Wacht. Ben je zeker dat ze er niet is? Check eerst even. \\
        & \textit{Wait. Are you sure that she is not there? Just check first.}\\
        \bottomrule
        & \\
        & (1) Example 1: Spontaneous conversation from a soap series \\
        & \\
        & \\
        \toprule
        \multirow{2}{=}[-9pt]{\textbf{REF} (verbatim)} & We hadden gehoopt eigenlijk na die die tweede dosis we zagen dat er goeie bescherming was dat dat zou blijven duren. \\
        & \textit{We had hoped actually after that that second dose we saw that there was good protection that that would last.} \\
        \midrule[0.01pt]
        \multirow{2}{=}{\textbf{REF} (subtitle)} & Wij hadden gehoopt dat het na de tweede doses zou blijven duren. \\
        & \textit{We had hoped that it would last after the second doses.} \\
        \midrule[0.01pt]
        \multirow{2}{=}[-9pt]{\textbf{HYP} (verbatim)} & En we hadden gehoopt eigenlijk na die die tweede dosis. We zagen dat de goeie bescherming was dat dat zou blijven duren.\\
        & \textit{And we had hoped actually after that that second dose. We saw that the good protection was that that would last.}\\
        \midrule[0.01pt]
        \multirow{2}{=}{\textbf{HYP} (subtitle)} & We hadden gehoopt na die tweede dosis dat dat zou blijven duren. \\
        & \textit{We had hoped after that second dose that it would last.}\\
        \bottomrule
        & \\
        & (2) Example 2: Live conversation with a virologist in a talk show \\
    \end{tabularx}
    \caption{Comparison of the reference (REF) verbatim and subtitle transcriptions to the hypotheses (HYP) generated by the outputs of the verbatim ASR decoder and the subtitle decoder. Translation in English is added for the reader in italics.}
    \label{tab:str}
\end{table*}

The examples in Table~\ref{tab:str} show several of the effects described in Section~\ref{sec:translation}. In Example 1, the colloquial pronoun ``\textit{ge}", common in spoken language, is translated to ``\textit{je}" in the subtitle. The expression ``\textit{zijt de}" (``are you") is a dialectal form of ``\textit{zijt ge}", which is generally written as ``\textit{ben je}". Notice that it is tagged with ${<}{*}d{>}$, and these tags are also predicted by the verbatim decoder. The colloquial word ``\textit{efkes}" (``shortly") is translated to it's standard version ``\textit{even}". Furthermore, the subtitle output tends to rephrase the spoken sentence towards standard written text, but retains the primary content of the message. Repetitions and speaking errors, such as the repeated ``\textit{die}" (``this") and ``\textit{dat}" (``that") in Example 2, are corrected in the subtitle. It is clear from the examples that the subtitle annotations are generally shorter than the verbatim annotations, and in many cases uninformative words and/or clauses are dropped.

\subsection{Generating Subtitles from ASR Outputs with LLMs} \label{sec:app-llm}
\vspace{0.2cm}
\noindent The previous section has highlighted the differences between verbatim transcripts that correspond to spoken language and subtitles that correspond to written language. The proposed methods in this paper are able to produce both annotation types directly from audio. In this section, we evaluate whether LLMs are capable to perform the same conversion task, but in the textual domain. More specifically, we task an LLM to convert a verbatim prediction of an ASR model to a subtitle using a specific prompt\footnote{The following prompt was found to work best with the LLM: "You have to create a subtitle based on an automatically generated verbatim transcription from an ASR model. To this end, you have to convert the spoken language (in the verbatim transcription) to standard written language, without changing the content. This involves correcting errors and spelling mistakes, and removing disfluencies and hesitations. The transcription and subtitle will be in Dutch. Only respond with the subtitle." More extensive prompts (e.g. as in \cite{liu2024recordingeyesechoingears}) with extensive requirements did not achieve better performance.}. 

First, we evaluate the zero-shot capabilities of the LLM using this prompt and without giving any reference examples. Additionally, we evaluate the LLM when giving a set of reference examples in the prompt. To this end, $N$ pairs of ASR hypotheses and reference subtitles are added to the prompt, with $N$ the number of examples (5, 20 or 50). The examples are derived from a list of randomly selected utterances from the \textit{subs-valid-14kh} test set. The ASR hypotheses in the examples correspond to the respective ASR model for which the outputs are used. 

We use the recent multilingual instruction-tuned Llama-3.1 model\footnote{\url{https://huggingface.co/meta-llama/Llama-3.1-8B-Instruct}} (8B parameters) \cite{dubey2024llama3herdmodels}, as well as a Dutch-finetuned language model based on Mistral-7B-v0.1 (7B parameters) \cite{jiang2023mistral7b} referred to as Geitje\footnote{\url{https://huggingface.co/BramVanroy/GEITje-7B-ultra}}. For Geitje, the prompt is translated to Dutch. 

Table~\ref{tab:subsfinal} depicts the BLEU scores on the long-form \textit{subs-valid-14kh} test set. The ASR models correspond to the previous section (Table~\ref{tab:asrfinal}). The BLEU scores are calculated for the ASR outputs as well as for the LLM generated subtitles with these ASR outputs. For our proposed method, we evaluate both the verbatim prediction and the subtitle prediction of the model as inputs for the LLM and analyse which one is most suitable.

\begin{table}
\begin{center}
\caption{BLEU scores (\%) on \textit{subs-valid-14kh} of subtitle prediction experiments ($\uparrow$) with integration of LLMs and comparison to Whisper. The LLMs are prompted zero-shot without any examples or with $N$ (5, 20 or 50) examples. For the proposed model with cascaded encoders using dual features, either the verbatim predictions or the subtitle predictions are used. The overall best model result is in bold. The best result for every model separately is underlined. \\}
\label{tab:subsfinal}
\resizebox{\columnwidth}{!}{
\begin{tabular}{@{\extracolsep{2pt}} l | c | c c c c | c c c @{}}
\toprule
\hfill \textbf{\textit{LLM}} & None & \multicolumn{4}{c}{\textbf{Llama-3.1}} & \multicolumn{3}{c}{\textbf{Geitje}} \\
\textbf{Model} \hfill \textit{\# Examples} & / & N=0~ & N=5~ & N=20~ & N=50~ & N=0~ & N=5~ & N=20~ \\
\midrule
E2E ASR & \underline{34.06}~ & 28.77~ & 33.35~ & 33.74~ & 33.94~ & 27.03~ & 29.63~ & 25.87~ \\
Proposed Model - Verbatim & 43.05~ & 36.05~ & 40.86~ & \underline{43.14}~ & 42.55~ & 34.81~ & 36.60~ & 34.39~ \\
Proposed Model (XL) - Verbatim & 44.35~ & 36.72~ & 42.19~ & \underline{44.68}~ & 43.60~ & 35.00~ & 37.78~ & 35.31~ \\
Proposed Model - Subtitle & \underline{59.40}~ & 38.24~ & 47.89~ & 50.52~ & 48.49~ & 48.65~ & 51.52~ & 41.47~ \\
Proposed Model (XL) - Subtitle & \underline{\textbf{61.76}}~ & 39.67~ & 49.71~ & 52.79~ & 50.76~ & 51.61~ & 53.72~ & 45.23~ \\
Whisper & \underline{50.58}~ & 36.19~ & 43.89~ & 46.89~ & 43.78~ & 43.27~ & 43.48~ & 36.37~ \\
Whisper finetuned & 39.57~ & 33.07~ & 40.36~ & \underline{41.88}~ & 41.47~ & 37.58~ & 35.07~ & 33.80~ \\
\bottomrule 
\end{tabular}
}
\end{center}
\end{table}

As the results in Table~\ref{tab:subsfinal} reveal, converting the verbatim transcript to a subtitle (in Dutch) is not a trivial task for an LLM, despite their impressive performance on many language-related tasks. The Dutch-finetuned language model Geitje seems to work better than the Llama model for zero-shot (without examples) transcript to subtitle conversion. However, the resulting BLEU scores of the LLM pipeline are still below the BLEU scores attained with the subtitle output of our proposed model. In fact, using an LLM in a zero-shot fashion even leads to a lower BLEU score than the verbatim outputs without LLM. When the LLM is given some examples of how to do the subtitling task, i.e. leveraging in-context learning, its predictions tend to be better subtitles than the original verbatim outputs. The amount of in-context learning is limited by the context window and attentive capabilities of the specific LLM. As a result, our method, which learns to generate subtitles based on both verbatim and acoustic content, has significant benefits. We remark that the performance of the LLM pipeline could be further improved by conditioning on longer historical contexts and by integrating additional contextual information \cite{liu2024recordingeyesechoingears}. Finetuning the LLM with paired subtitle-verbatim examples or providing more accurate descriptions of the actual subtitling protocol (which we don't have) might yield further improvements, although adapting to verbatim transcripts is probably necessary.

Moreover, we note that the BLEU score for the standard Whisper model is already quite good. In combination with the results in Table~\ref{tab:asrfinal}, this further indicates that Whisper transcriptions are closer to standard written language than they are to verbatim transcriptions. In e.g. assessment tasks such as fluency measurement, this is a downside resulting from not distinguishing transcript types during weakly supervised training, which we aim to solve with the methods in this paper. If Whisper is finetuned on verbatim data, its subtitling capabilities are greatly reduced. In any case, our proposed models with dedicated subtitle decoders still attains the highest BLEU scores, although it should be noted that they are aware of the subtitling protocol as they were trained on subtitles from the same broadcaster.

\end{document}